\documentclass[%
 aip,
 amsmath,amssymb,
 reprint,
 floatfix,
]{revtex4-1}

\usepackage{color}
\usepackage[usenames,dvipsnames,svgnames,table]{xcolor}
\usepackage[colorlinks=true,linkcolor=blue,urlcolor=blue,citecolor=blue]{hyperref}

\usepackage[normalem]{ulem}
\usepackage{mathtools}
\usepackage{graphicx}
\usepackage{dcolumn}
\usepackage{array}
\usepackage{lipsum}
\usepackage{bm}
\usepackage{subfigure}
\usepackage{amssymb}
\usepackage{multirow}
\usepackage{tabularx}
\usepackage{amsmath}
\usepackage{braket}
\usepackage{csquotes}
\graphicspath{{plots/}}
 \usepackage{lipsum}
\usepackage{mathrsfs}
\usepackage{MnSymbol}

\newcommand{\beq}{\begin{equation}}
\newcommand{\eeq}{\end{equation}}
\newcommand{\bea}{\begin{eqnarray}}
\newcommand{\eea}{\end{eqnarray}}

\makeatletter
\def\@email#1#2{%
 \endgroup
 \patchcmd{\titleblock@produce}
  {\frontmatter@RRAPformat}
  {\frontmatter@RRAPformat{\produce@RRAP{*#1\href{mailto:#2}{#2}}}\frontmatter@RRAPformat}
  {}{}
}%
\makeatother

\begin{document}

\title{Effects of Mosaic Crystal Instrument Functions on X-ray Thomson Scattering Diagnostics}

\author{Thomas~Gawne}
\email{t.gawne@hzdr.de}
\affiliation{Center for Advanced Systems Understanding (CASUS), D-02826 G\"orlitz, Germany}
\affiliation{Helmholtz-Zentrum Dresden-Rossendorf (HZDR), D-01328 Dresden, Germany}

\author{Hannah~Bellenbaum}
\affiliation{Center for Advanced Systems Understanding (CASUS), D-02826 G\"orlitz, Germany}
\affiliation{Helmholtz-Zentrum Dresden-Rossendorf (HZDR), D-01328 Dresden, Germany}
\affiliation{Institut f\"ur Physik, Universit\"at Rostock, D-18057 Rostock, Germany}

\author{Luke~B.~Fletcher}
\affiliation{SLAC National Accelerator Laboratory, Menlo Park California 94025, USA}

\author{Karen~Appel}
\affiliation{European XFEL, D-22869 Schenefeld, Germany}

\author{Carsten~Baehtz}
\affiliation{Helmholtz-Zentrum Dresden-Rossendorf (HZDR), D-01328 Dresden, Germany}

\author{Victorien~Bouffetier}
\affiliation{European XFEL, D-22869 Schenefeld, Germany}

\author{Erik~Brambrink}
\affiliation{European XFEL, D-22869 Schenefeld, Germany}

\author{Danielle~Brown}
\affiliation{SLAC National Accelerator Laboratory, Menlo Park California 94025, USA}

\author{Attila~Cangi}
\affiliation{Center for Advanced Systems Understanding (CASUS), D-02826 G\"orlitz, Germany}
\affiliation{Helmholtz-Zentrum Dresden-Rossendorf (HZDR), D-01328 Dresden, Germany}

\author{Adrien~Descamps}
\affiliation{School of Mathematics and Physics, Queens University Belfast, 
Belfast BT7 1NN, United Kingdom}

\author{Sebastian~Goede}
\affiliation{European XFEL, D-22869 Schenefeld, Germany}

\author{Nicholas~J.~Hartley}
\affiliation{SLAC National Accelerator Laboratory, Menlo Park California 94025, USA}

\author{Marie-Luise~Herbert}
\affiliation{Institut f\"ur Physik, Universit\"at Rostock, D-18057 Rostock, Germany}

\author{Philipp~Hesselbach}
\affiliation{GSI Helmholtz Centre for Heavy Ion Research, 
D-64291 Darmstadt, Germany}
\affiliation{Institut f\"ur Angewandte Physik, Goethe-Universit\"at, D-60438 Frankfurt am Main, Germany}

\author{Hauke~H\"oppner}
\affiliation{Helmholtz-Zentrum Dresden-Rossendorf (HZDR), D-01328 Dresden, Germany}

\author{Oliver~S.~Humphries}
\affiliation{European XFEL, D-22869 Schenefeld, Germany}

\author{Zuzana~Kon\^opkov\'a}
\affiliation{European XFEL, D-22869 Schenefeld, Germany}

\author{Alejandro~Laso~Garcia}
\affiliation{Helmholtz-Zentrum Dresden-Rossendorf (HZDR), D-01328 Dresden, Germany}

\author{Bj\"orn~Lindqvist}
\affiliation{Institut f\"ur Physik, Universit\"at Rostock, D-18057 Rostock, Germany}

\author{Julian~L\"utgert}
\affiliation{Institut f\"ur Physik, Universit\"at Rostock, D-18057 Rostock, Germany}

\author{Michael~J.~MacDonald}
\affiliation{Lawrence Livermore National Laboratory (LLNL), California 94550~Livermore,~USA}

\author{Mikako~Makita}
\affiliation{European XFEL, D-22869 Schenefeld, Germany}

\author{Willow~Martin}
\affiliation{SLAC National Accelerator Laboratory, Menlo Park California 94025, USA}

\author{Mikhail~Mishchenko}
\affiliation{European XFEL, D-22869 Schenefeld, Germany}

\author{Zhandos~A.~Moldabekov}
\affiliation{Center for Advanced Systems Understanding (CASUS), D-02826 G\"orlitz, Germany}
\affiliation{Helmholtz-Zentrum Dresden-Rossendorf (HZDR), D-01328 Dresden, Germany}

\author{Motoaki~Nakatsutsumi}
\affiliation{European XFEL, D-22869 Schenefeld, Germany}

\author{Jean-Paul~Naedler}
\affiliation{Institut f\"ur Physik, Universit\"at Rostock, D-18057 Rostock, Germany}

\author{Paul~Neumayer}
\affiliation{GSI Helmholtz Centre for Heavy Ion Research, 
D-64291 Darmstadt, Germany}

\author{Alexander~Pelka}
\affiliation{Helmholtz-Zentrum Dresden-Rossendorf (HZDR), D-01328 Dresden, Germany}

\author{Chongbing~Qu}
\affiliation{Institut f\"ur Physik, Universit\"at Rostock, D-18057 Rostock, Germany}

\author{Lisa~Randolph}
\affiliation{European XFEL, D-22869 Schenefeld, Germany}

\author{Johannes Rips}
\affiliation{Institut f\"ur Physik, Universit\"at Rostock, D-18057 Rostock, Germany}

\author{Toma~Toncian}
\affiliation{Helmholtz-Zentrum Dresden-Rossendorf (HZDR), D-01328 Dresden, Germany}

\author{Jan~Vorberger}
\affiliation{Helmholtz-Zentrum Dresden-Rossendorf (HZDR), D-01328 Dresden, Germany}

\author{Lennart~Wollenweber}
\affiliation{European XFEL, D-22869 Schenefeld, Germany}

\author{Ulf~Zastrau}
\affiliation{European XFEL, D-22869 Schenefeld, Germany}

\author{Dominik~Kraus}
\affiliation{Institut f\"ur Physik, Universit\"at Rostock, D-18057 Rostock, Germany}
\affiliation{Helmholtz-Zentrum Dresden-Rossendorf (HZDR), D-01328 Dresden, Germany}

\author{Thomas~R.~Preston}
\affiliation{European XFEL, D-22869 Schenefeld, Germany}

\author{Tobias~Dornheim}
\affiliation{Center for Advanced Systems Understanding (CASUS), D-02826 G\"orlitz, Germany}
\affiliation{Helmholtz-Zentrum Dresden-Rossendorf (HZDR), D-01328 Dresden, Germany}

\begin{abstract}
Mosaic crystals, with their high integrated reflectivities, are widely-employed in spectrometers used to diagnose high energy density systems. X-ray Thomson scattering (XRTS) has emerged as a powerful diagnostic tool of these systems, providing in principle direct access to important properties such as the temperature via detailed balance. However, the measured XRTS spectrum is broadened by the spectrometer instrument function (IF), and without careful consideration of the IF one risks misdiagnosing system conditions. Here, we consider in detail the IF of 40~$\mu$m and 100~$\mu$m mosaic HAPG crystals, and how the broadening varies across the spectrometer in an energy range of 6.7-8.6~keV. Notably, we find a strong asymmetry in the shape of the IF towards higher energies.
As an example, we consider the effect of the asymmetry in the IF on the temperature inferred via XRTS for simulated 80~eV CH plasmas, and find that the temperature can be overestimated if an approximate symmetric IF is used.
We therefore expect a detailed consideration of the full IF will have an important impact on system properties inferred via XRTS in both forward modelling and model-free approaches.

\end{abstract}

\maketitle

\section{Introduction\label{sec:introduction}}

Warm dense matter (WDM) is ubiquitous in the universe: it is prevalent in astrophysical environments such as Jovian, solar and white dwarf objects~\cite{guillot1999interiors,Bailey2015-hi,Kritcher2020}, and it is routinely produced in the laboratory at laser~\cite{hoarty2013observations} and free electron laser (FEL)~\cite{Vinko2012-fc} facilities. It is therefore a state of matter that boasts huge physical interest, particularly as its practical applications in inertial confinement fusion~\cite{Zylstra2022-kp,LawsonCriterion-2022,Gain-2024} and the development of new materials~\cite{Kraus2016-hc,Miao2020-ew} continues to mature.
However, the rigorous diagnosis of conditions in WDM produced in experiments remains a persistent challenge. The extreme conditions that characterise WDM -- high temperatures, pressures, and densities -- are highly transient and destructive,  requiring the need for $\textit{in-situ}$ diagnosis.

A multitude of techniques have been developed to infer WDM conditions based on the interaction of x-rays with the target. In that respect, x-ray Thomson scattering (XRTS) has long been a workhorse of plasma diagnostics~\cite{Glenzer_RMP_2009}. By probing the electronic dynamic structure factor (DSF) of the system, XRTS measurements contain the full details of the plasma temperature, density, ionization and electronic correlations of the system, if one knows how to extract this information. Typically, the XRTS intensity $I({\bm q}, E)$ is expressed as the convolution of the DSF $S({\bm q}, \omega)$ with the source-and-instrument function (SIF) $\xi(E_0)$ of the experiment~\cite{Glenzer_RMP_2009,Dornheim_T2_2022}:
\begin{equation}
    \label{eq:XRTS}
    I({\bm q}, E) = (S \circledast \xi)({\bm q}, E) \, ,
\end{equation}
where ${\bm q}$ is the scattering vector, $E$ is the energy of the scattered photon measured on the spectrometer, $E_0$ is the energy of the incident photon, and $\hbar \omega = E_0 - E$ is the energy loss.
The SIF contains all the effects that broaden the measured XRTS spectrum, specifically the spectrum of the incident probe beam (the source) and the instrument function of the spectrometer.
The temperature $k_BT = \beta^{-1}$ of a system in thermal equilibrium, for example, may in principle be extracted using the detailed balance relation of the DSF:
\begin{equation}
    S({\bm q}, -\omega) = e^{-\beta \hbar \omega} S({\bm q}, \omega) \, ,
\end{equation}
while other properties may be derived from the frequency moments of the DSF~\cite{giuliani2008quantum,Crowley_2013}.
Of course, in order to access these properties, one needs to remove the broadening by the SIF. Since numerical deconvolution is unstable due to the finite spectral windows and noise, so-called forward modelling has emerged as the standard approach for interpreting XRTS data. In this approach, the XRTS spectrum is fitted using a model of the DSF that has been broadened by the SIF, with the various plasma parameters being optimized to achieve a best fit. Of course, this leads to the inferred conditions being dependent on the chosen DSF model, such as the decomposition of the system into effectively bound and free electronic populations within the popular Chihara approach~\cite{Chihara_JoP_1987,Chihara_JoP_2000,Gregori_PRE_2003}. More recent developments have proposed to deconvolve the DSF using the two-sided Laplace transform and represent the DSF in the imaginary time ($\tau$) domain as the imaginary time correlation function (ITCF) $\mathcal{F}(\bm{q}, \tau)$~\cite{Dornheim_MRE_2023}:
\begin{equation}
    \label{eq:ITCF}
    \mathcal{F}(\bm{q}, \tau) \equiv \mathcal{L}[S](\bm{q}, \tau)  = \frac{\mathcal{L}[S\circledast\xi]}{\mathcal{L}[\xi]}  = \frac{\mathcal{L}[I]}{\mathcal{L}[\xi]} \, .
\end{equation}
By definition, the ITCF contains the exact same information as the DSF, so it can also be used to infer system conditions -- for example, detailed balance is expressed by the symmetry of the ITCF around $\tau = \beta/2$~\cite{Dornheim2022-fh}:
\begin{equation}
    \mathcal{F}(\bm{q}, \tau) = \mathcal{F}(\bm{q}, \beta - \tau) \, .
\end{equation}
Since this approach does not require an input model for the DSF, the ITCF method provides an in-principle model-free interpretation of XRTS experiments for arbitrarily complex materials and mixtures.
However, while some contributions to the SIF can be directly measured in experiments (such as the beam profile), the spectrometer IF is typically not measured and is instead approximated by some model. The chosen model for the SIF will affect the results inferred from both the ITCF approach and forward modelling.

An x-ray spectrometer typically consists of two components: a dispersive crystal and a photosensitive detection method (diode, scintillator, detector, etc). As the light emanating
from a target is polychromatic,
in order to measure the spectrum the different photon energies need to be separated from each other. This is often done using a crystal with an appropriate lattice spacing to disperse the x-rays in angle, which are then detected spatially offset in the detection plane. The spatial position of a photon on the detector can be related to its energy via Bragg's law and the geometry of the spectrometer~\cite{vonHamos_1932,zachariasen1994theory}.

Mosaic crystals are a popular choice for the dispersive crystal. While perfect crystals consist of uniform layers of atoms, mosaic crystals consist of small perfect crystallites 
with normal vectors randomly distributed about the surface normal~\cite{zachariasen1994theory,FREUND1988461,legall2006new,Gerlach_JAC_2015}. As a result, in contrast to perfect crystals where the Bragg condition will only be satisfied in a certain region in the crystal depending on the rocking curve, on a mosaic crystal it can be satisfied anywhere on its surface. The solid angle coverage of a mosaic crystal is therefore substantially higher than that of a perfect crystal, resulting in a much higher integrated reflectivity~\cite{Legall_High_2006,2009_Legall_Efficient}. In the diagnostics of WDM systems in the laboratory, where these systems are often transient, this higher reflectivity has led to the widespread adoption of mosaic crystals as they collect more photons in a given shot.
But this increased reflectivity comes at the cost of spectral resolution as the distribution of the crystallites also broadens the measured spectrum. For example, in widely-used Highly Annealed Pyrolytic Graphite (HAPG) crystals, the distribution of crystallites is approximately Lorentzian~\cite{Zastrau_JoI_2013,Gerlach_JAC_2015}, and therefore the crystal broadening of spectra has often been treated as a convolution of the source spectrum with a Voigt profile~\cite{Preston_JoI_2020,MacDonald_PoP_2021}.

However, recent ray tracing simulations~\cite{2012_Zastrau_Focal,Smid_CPC_2021} and experimental data~\cite{Pak_RoSI_2004,Doeppner_RoSI_2008,Voigt_Ramen_2021} indicate that the actual broadening by mosaic crystals is very asymmetric towards higher energies.
An asymmetric SIF would have consequences on conditions derived from XRTS measurements if an approximate SIF that is symmetric -- or one without sufficient asymmetry -- were used in its place. A clear example would be that the temperature extracted via detailed balance would be incorrect as the upshifted side of the DSF would be inferred to be larger to compensate for the increase in intensity by the SIF.

On a more basic level, a further question also arises as to the validity of the assumption that the SIF is applied to the DSF as a convolution. As the interaction of x-rays with materials depends on the photon energy of the x-ray~\cite{2009_Legall_Efficient,HENKE19821}, the spectrometer instrument function (IF) must depend on the x-ray photon energy as a parameter and not simply on the energy loss:
\begin{equation}
    \label{eq:Kernel}
    I({\bm q}, E) = \int_{-\infty}^\infty \xi(E-E_0; E_0) \int_{-\infty}^\infty S({\bm q}, \omega) B(E_0-\hbar\omega) \,  d\omega d E_0  \, ,
\end{equation}
where the broadening by the probe beam profile $B(E)$ has been separated from the spectrometer IF $\xi(E-E_0; E_0)$.
In general then, the removal of the SIF will be a kernel problem that will be extremely challenging to invert, unless the photon energy parameterization is sufficiently weak within the spectral range that Eq.~(\ref{eq:Kernel}) may be approximated by a convolution.

In this paper, we present a systematic analysis of the broadening by mosaic crystals and the effect it has on plasma conditions inferred in XRTS measurements.
To do so, we use Schlesiger \textit{et al.}'s multi-reflection mosaic crystal model~\cite{Schlesiger_JAC_2017} to calculate the differential reflectivity along paths a photon can take from the source to the detector.
The model is found to have good agreement with experimental measurements taken at the European XFEL on ambient Al and polypropylene. Notably, we find the mosaic broadening and depth broadening result in highly asymmetric broadening towards the higher energies. The instrument function is also found to be extremely extended, with substantial intensity even more than 100 eV away from the central photon energy.
Furthermore, it is demonstrated that such strong asymmetry in the beam
has implications on conditions inferred from XRTS via forward modelling and the model-free method if an approximate symmetric SIF is used.
Finally, we investigate the validity of the approximation of the SIF as a convolution problem rather than a kernel problem.

The paper is organised as follows: in Section~\ref{sec:Model} we introduce the model for the mosaic crystal instrument function; in Section~\ref{sec:results} we compare the model to experimental data, and examine the effects of the IF on inferred experimental conditions; and finally in Section~\ref{sec:summary} we provide concluding remarks.

\section{Broadening by Mosaic Crystals\label{sec:Model}}

\begin{figure}
    \centering
    \includegraphics[width=\columnwidth,keepaspectratio]{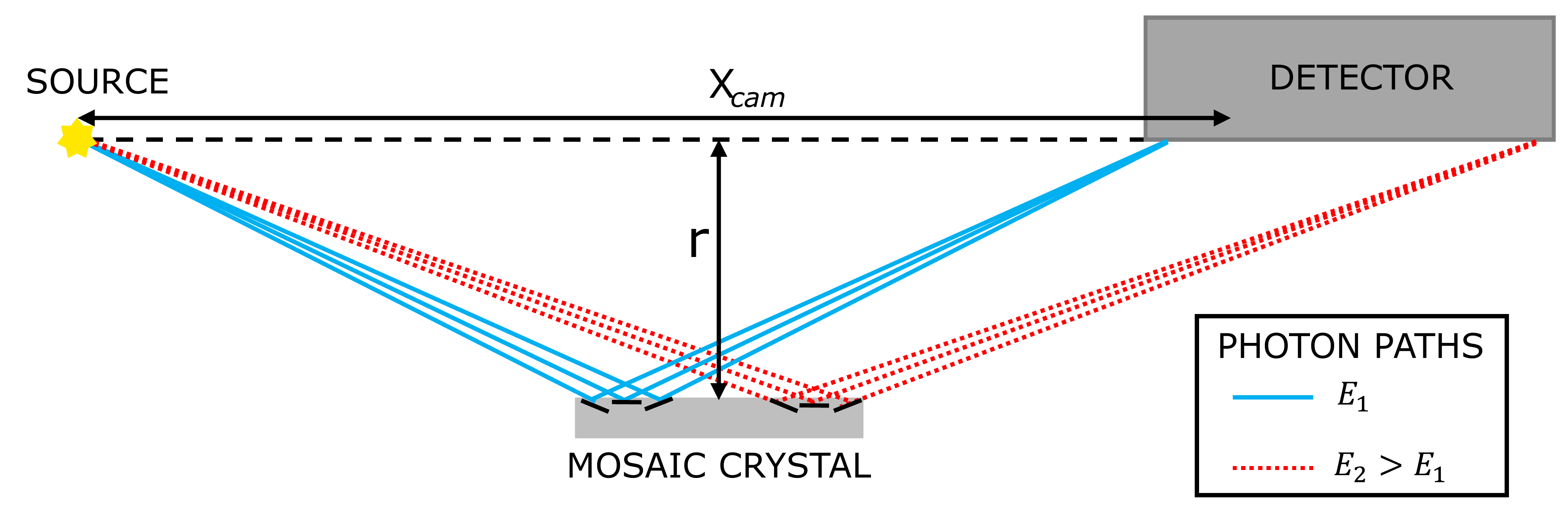}
    \caption{A two-dimensional schematic of a von H\'amos detector with a mosaic crystal, with the paths of two different photon energies, $E_1$ and $E_2>E_1$, to the detector also shown. In full von H\'amos geometry, the crystal is cylindrically bent with a radius of curvature $r$, while $X_{\rm cam}$ is the distance from the source to a point on the detector. Mosaic focusing in this geometry is represented by photons incident on the crystal at different angles reflecting off crystallites at angles to the surface normal, which then focus back to the same point on the detector.
    In the present two-dimensional geometry, the crystal and detector are lines along the dispersion axis, with the crystal having a finite thickness.
    } 
    \label{fig:VonHamos}
\end{figure}

To start, we consider some of the important contributions to the crystal function, namely the depth broadening, the mosaic broadening, and the intrinsic rocking curve (IRC). Note that the first and last contributions are universal to (finite thickness) crystals. We also describe here the model used to estimate the crystal broadening.

For simplicity and practicality, we consider a spectrometer in von H\'amos geometry~\cite{vonHamos_1932} (see Fig.~\ref{fig:VonHamos}). In this geometry, a cylindrically bent crystal is used to focus the dispersed x-rays on to a single dispersion axis on the detector. By collecting all the scattered photons on to a single line, it is easier to observe events above the detector noise.
As the bending process severely degrades the resolution of perfect crystals anyway, mosaic crystals are often used in von H\'amos spectrometers to further increase the efficiency of the detector by taking advantage of their high reflectivity, and because the geometry allows one to utilise mosaic focusing~\cite{Preston_JoI_2020}, as shown in Fig.~\ref{fig:VonHamos}.
Due to their very high collection efficiency, mosaic crystal von H\'amos spectrometers are widely-employed at FEL and laser facilities; therefore, the choice of von H\'amos geometry here is relevant for typical applications of mosaic crystals, though we expect the results in this work can be generalised to alternative geometries.
In von H\'amos geometry, the distance from the source to a point on the detector $X_{\rm cam}$ can be readily converted to a photon energy using the dispersion equation~\cite{vonHamos_1932}:
\begin{equation}
    \label{eq:vonHamos}
    E_{\rm cam} = \frac{hc}{2d} \sqrt{1 + \frac{X_{\rm cam}^2}{4r^2}} \, ,
\end{equation}
where $d$ is the crystal lattice spacing, $r$ is the distance of the crystal below the source-detector axis, $h$ is Planck's constant, and $c$ is the speed of light.

For further simplicity and efficiency, we only treat the spectrometer in two-dimensions, as is shown in Fig.~\ref{fig:VonHamos}. In other words, the source is treated as a point, the detector and the crystal are treated as lines along the dispersion axis, and the crystal has a finite thickness.
We note this two-dimensional setup inherently does not explicitly treat any broadening effects off the dispersion axis on the detector and along the bent crystal.
Nevertheless, as will be demonstrated, this simplified treatment of the spectrometer still provides good agreement with experimental measurements.

While we focus here on the effects of the crystal broadening, some mention should be given to the effect of the detection method on the measured spectrum. As an area detector has a finite size, its most immediate effect is to force the measurement to a finite spectral window.
The continuous photon energy distribution is discretized into the finite-size pixels -- usually, though, the pixel size is small enough that the produced spectrum will look smooth and continuous.
 By using the dispersion relation of the spectrometer, the spectrum on the detector represents number of photons detected at a specific photon energy. The pixels also subtend a finite solid angle, which can also be readily accounted for, though typically the broadening introduced by this is much less than that intrinsically from bent crystals.
In short, while the detector does form part of the spectrometer IF, its effects are generally well-understood.

Finally, to clarify the terminology used in this section, note that we distinguish the crystal rocking curve and the IRC: the former accounts for all effects (including the mosaicity and the IRC) that allows for scattering to occur off the Bragg condition, while the latter refers only to the rocking curve of a perfect crystal~\cite{zachariasen1994theory}.

\subsection{Depth Broadening}
\begin{figure*}
    \centering
    \includegraphics[width=\textwidth,keepaspectratio]{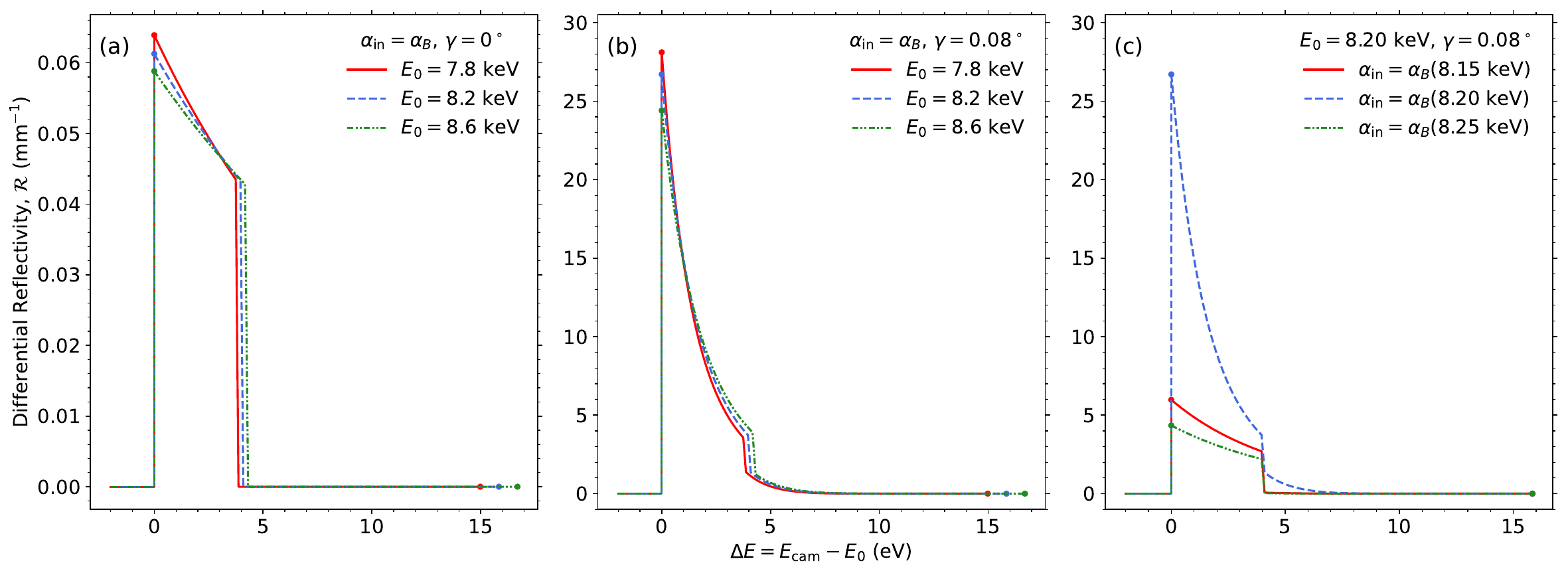}
    \caption{The differential reflectivity from depth broadening in three scenarios involving a 40~$\mu$m thick graphite crystal, considering up to $p=4$ reflections. Note the marked increase in reflectivity when mosaicity is taken into account.
    In all cases, the circular markers indicate the energy difference on detector that a photon can be reflected to.
    (a) Depth broadening for a photon with different energies incident at its Bragg angle on a crystal without mosaicity.
    Note that the shift on the detector is only for $\Delta E > 0$, and that the maximum $\Delta E$ is different for each photon energy. 
    (b) The same as (a), but now the crystal has a mosaicity $\gamma = 0.08^\circ$.
    (c) Depth broadening for a fixed photon energy on a mosaic crystal at different angles of incidence.
    } 
    \label{fig:DepthEffects}
\end{figure*}

Depth broadening is simply the broadening that results from the photon travelling into a finite depth of the crystal before scattering. The reflection event therefore occurs at a distance slightly further away from the source than when reflecting off the surface, so the photon will be incident on the detector slightly further away from the source. From Eq.~(\ref{eq:vonHamos}), the depth broadening is therefore towards higher energies. As the finite thickness of a crystal provides more opportunities for a photon to scatter off a crystal layer, thicker crystals will have a higher integrated reflectivity at the cost of resolution.

In the context of mosaic crystals, Schlesiger \textit{et al.}~\cite{Schlesiger_JAC_2017} described a model for multiple reflections within the crystal; that is, the photon scatters off the crystal layers multiple times before exiting the crystal. In this model, which we will briefly outline here, the differential reflectivity for a reflection of order $p$ is given by the equation:
\begin{equation}
\begin{aligned}
    \label{eq:DiffRefl}
    \mathcal{R}_p(\alpha_{\rm in}, z_{\rm eff}, \gamma_C, T_c, E_0) &= G_p^{Tc}(z_{\rm eff}) \sigma^{2p-1}  \exp({-\mu_{\rm eff} z_{\rm eff}}) \, ,
\end{aligned}
\end{equation}
where an odd ($2p-1$) number of reflections is required for the photon to leave the front surface of the crystal and be detected. In Eq.~(\ref{eq:DiffRefl}), $\alpha_{\rm in}$ is the angle of incidence of the photon on the crystal, $z_{\rm eff}$ is the effective depth a photon travels through the crystal, $T_c$ is the thickness of the crystal, $\gamma_C$ is the full width at half-maximum (FWHM) of the rocking curve of the crystal, and $E_0$ is the energy of the incident photon. $G_p^{Tc}(z_{\rm eff})$ is a degeneracy factor that accounts for the number of paths that produce the same $z_{\rm eff}$.
The reflecting power $\sigma$ of a crystal or crystallite with thickness $dT$ is given by~\cite{zachariasen1994theory}:
\begin{equation}
    \sigma dT = \left| \frac{e^2 F_H}{m_e c^2 V} \right|^2 \lambda^3 \frac{1 + \cos^2(2\alpha_B)}{2 \sin(2\alpha_B) \sin(\alpha_B)}  W(\Delta \alpha, \gamma_C) dT \, ,
\end{equation}
where $F_H$ is the structure factor for the Miller indices $H$, $V$ is the volume of the unit cell, $\lambda$ is the wavelength of the incident photon, $\alpha_B$ is the Bragg angle for the photon energy $E_0$, $W(\Delta \alpha, \gamma_C)$ is the rocking curve function, and $\Delta \alpha = \alpha_B - \alpha_{\rm in}$.
Lastly, the term $\mu_{\rm eff}$ is the effective attenuation coefficient~\cite{Schlesiger_JAC_2017}:
\begin{equation}
    \mu_{\rm eff} = \mu_0 \left( \frac{1}{\sin(\alpha_{\rm in})} + \frac{1}{\sin(\alpha_{\rm out})} \right) + 2\sigma \, ,
\end{equation}
where $\mu_0$ is the attenuation coefficient of a photon with energy $E_0$ into the material, and $\alpha_{\rm out} = 2\alpha_B - \alpha_{\rm in}$ is the angle of the outgoing diffracted ray.

The sum of the differential reflectivity in Eq.~(\ref{eq:DiffRefl}) over the reflection orders $p$, $\mathcal{R}(\alpha_{\rm in}, z_{\rm eff}, \gamma, T_c, E_0)$, provides the probability a photon with an angle of incidence $\alpha_{\rm in}$ will travel an effective depth $z_{\rm eff}$ into the crystal. As these parameters also determine the path along which the photon leaves the crystal, it also provides the photon distribution at the detection plane.

For a crystal with no mosaicity, the differential reflectivity from the depth broadening in 40~$\mu$m graphite is plotted in Fig.~\ref{fig:DepthEffects}~(a).
For first order reflections of a monochromated beam, the depth broadening effect is essentially a convolution with an exponential that has hard cuts at distances associated with the front surface (first opportunity to reflect) and back surface (when the photon has transmitted through the crystal) of the crystal. The higher order reflections are present, though in practice these decay very quickly with increasing reflection order due to the $\sigma^{2p-1} \exp(-\mu_{\rm eff} z_{\rm eff})$ terms. Indeed, in Fig.~\ref{fig:DepthEffects}~(a) the only notable reflections are those of the first order.
It should also be noted that even in the absence of mosaicity, the depth broadening introduces asymmetry into the crystal instrument function as it only broadens to higher energies.

Even at just the level of depth broadening, for a polychromatic beam we can begin to observe some deviations of the broadening from a simple convolution. Notably, $\sigma$ depends directly on the photon energy (via $\lambda$); and $\sigma$ and $\mu_{\rm eff}$ depend on the Bragg angle, which depends on the photon energy too. Additionally, the angle of approach affects how far a photon needs to travel through the crystal before it is detected, and indeed how far it can travel before it reaches the back surface. These effects together explain the differences between the three photon energies plotted in Fig.~\ref{fig:DepthEffects}~(a). An overall conclusion that will be observed in this section is that, in general, broadening effects will depend on the photon energy and the geometry of the spectrometer.

\subsection{Mosaic Broadening}

\begin{figure*}
    \centering
    \includegraphics[width=\textwidth,keepaspectratio]{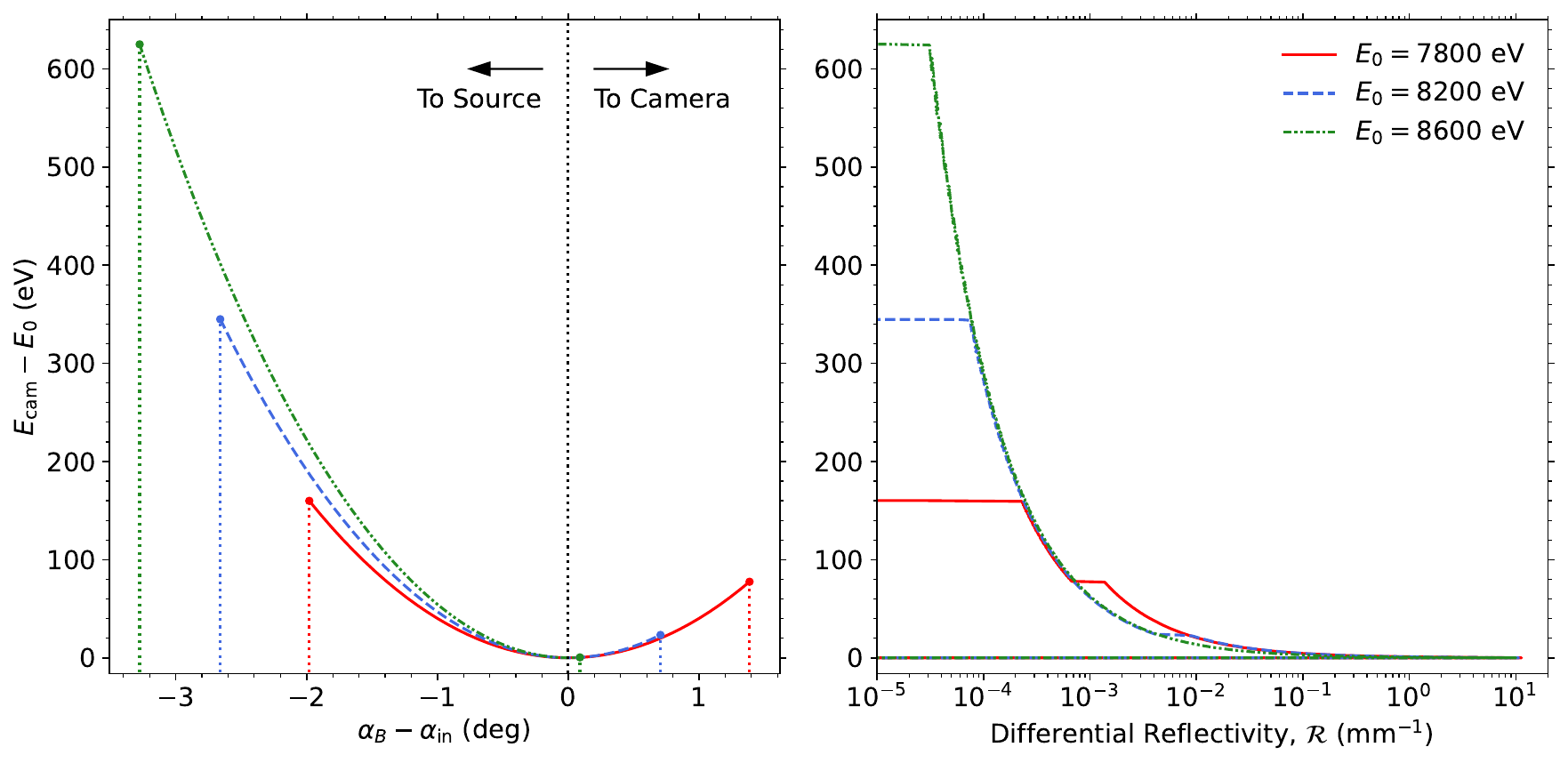}
    \caption{A break down of the effects contributing to mosaic broadening, just for reflections from the surface of a 80~mm long HAPG crystal. Left: A plot of Eq.~(\ref{eq:AngleToCam}) in von H\'amos geometry for different incident photon energies $E_0$, showing the energy shift on the detector versus the shift in the angle of incidence on the crystal from the photon's Bragg angle. The mosaic broadening is towards higher energies on the detector than the photon energy, and there is overlap in the position on the detector that the crystal will reflect to. The vertical dotted lines indicate the edges of the crystal.
    Right: The mosaic function as it appears on the detector. The drops in the reflectivity are when the edges of the crystal are reached.
    }
    \label{fig:MosaicOnCam}
\end{figure*}


The mosaic broadening is a result of the distribution of the crystallites to the surface normal.
In the differential reflectivity $\mathcal{R}_p$, the mosaicity of the crystal appears in the $W(\Delta \alpha, \gamma)$ term in $\sigma$, where $W$ is the angular distribution function of the crystallites; $\gamma$ is the mosaicity, typically denoting the full width at half maximum (FWHM) of the distribution; and $\Delta \alpha$ now also represents the angular deviation of a crystallite from the surface normal. This function can be measured by measuring the reflectivity at different angles of incidence along the crystal~\cite{Gerlach_JAC_2015}. For example, in HOPG the graphite crystallites have a Gaussian distribution~\cite{1996_Beckoff_SPIE,1996_Freund_SPIE}; while in HAPG they have a distribution that resembles the sum of two Lorentzians~\cite{Gerlach_JAC_2015}. For the purpose of the work presented here, we consider the HAPG crystallite distribution to be a single Lorentzian as this has been observed to be quite accurate in previous works~\cite{Zastrau_JoI_2013,Schlesiger_JAC_2017}, and it greatly simplifies the fitting procedure given the lack of actual measurements of the mosaic function in the relevant crystals.

As is usually the case, the higher integrated reflectivity of a mosaic crystal comes at the cost of a worse resolution as the angular distribution of the crystallites also broadens the measured spectrum -- while the Bragg angle is satisfied across the crystal, the point the photon arrives on the detector is still shifted from its nominal position. The mosaic broadening function has previously been considered to resemble the mosaic distribution function~\cite{Preston_JoI_2020}.
However, in von H\'amos geometry, the mosaicity actually only results in broadening towards higher energies, which can be demonstrated trigonometrically. In von H\'amos geometry, a scattered photon will hit the detector at a distance:
\begin{equation}
\begin{aligned}
    \label{eq:AngleToCam}
    X_{\rm cam} &= r \left[\cot(\alpha_{\rm in}) + \cot(\alpha_{\rm out})\right] \\
    &= r\left[\cot(\alpha_B-\Delta \alpha) + \cot(\alpha_B+\Delta \alpha)\right] \, ,
\end{aligned}
\end{equation}
away from the source. This equation is plotted in the left panel of Fig.~\ref{fig:MosaicOnCam}. When $\alpha_{\rm in} = \alpha_B$, the nominal position of the photon on the detector is obtained, $X_0= 2r\cot(\alpha_B)$.
Generally speaking, $\Delta \alpha$ will be very small across a crystal, so Eq.~(\ref{eq:AngleToCam}) can be expanded as a Taylor series:
\begin{equation}
\begin{aligned}
    X_{\rm cam} &=  r\left[ 2\cot(\alpha_B) + 2\cot(\alpha_B)\csc^2(\alpha_B) \Delta\alpha^2 + \mathcal{O}(\Delta\alpha^4) \right] \\
        &=  X_0 +  X_0\csc^2(\alpha_B) \Delta\alpha^2 + \mathcal{O}(\Delta\alpha^4) \, .
\end{aligned}
\end{equation}
From this quadratic, it is clear $X_{\rm cam} \ge X_0$, for all $\alpha_B$ and $\Delta \alpha$, which is what is observed in the left panel of Fig.~\ref{fig:MosaicOnCam}.

The relationship in Eq.~(\ref{eq:AngleToCam}) also implies that there are cut-offs in the spectrum when the edges of the crystal are reached as this function is not one-to-one with respect to the angle of incidence. This effect is plotted in the right panel of Fig.~\ref{fig:MosaicOnCam}, which shows this effect in HAPG across a realistic detector range ($E_{\rm cam} \sim 7800-8700$~eV). Evidently, there are quite pronounced differences in the reflectivity across the detector range due to these cut offs.

What is particularly noteworthy in Fig.~\ref{fig:MosaicOnCam} is that the energy broadening is extremely extended due to the Lorentzian nature of the HAPG crystallite distribution. While the distribution does cover a range of orders of magnitude, for sufficiently bright features such a large deviation would be detectable: as shall be demonstrated later, the quasi-elastic signal from a monochromated beam can still be meaningfully detected out to energies $>150$~eV away. So while the most apparent broadening from a mosaic crystal may appear relatively narrow (e.g. $\sim 2.5$~eV in HAPG~\cite{Preston_JoI_2020}), the tails of the spectrum are still important to consider as they do not decay quickly.

Another consequence of mosaic broadening is that it affects the shape of the depth broadening as well, in particular enhancing the prominence of the higher order reflections relative to the first order. An example is shown in the comparison of Fig.~\ref{fig:DepthEffects}~(a) and (b), where the overall shape of the depth broadening is dramatically changed with a finite mosaicity. Indeed, the position of the first order cut-off depends on the relative thickness and mosaicity of the crystal~\cite{Schlesiger_JAC_2017}. Also shown in Fig.~\ref{fig:DepthEffects}~(c) is the depth broadening when changing the angle of incidence away from the nominal Bragg angle. Due to the mosaicity, the differential reflectivity is still finite though reduced because of the crystallite distribution function.

Like the depth broadening, the mosaicity also depends on the incident photon energy though it is a quite weak dependence~\cite{Zastrau_JoI_2013,Gerlach_JAC_2015}. Overall, it is apparent that the mosaic broadening has quite a strong dependence on the spectrometer geometry, specifically the relative positions of the source, crystal, and detector. It is also clear that the mosaic and depth broadening only contribute to higher energies, meaning that the final crystal instrument function must be asymmetric towards higher photon energies.

\subsection{Intrinsic Rocking Curve}

\begin{figure}
    \centering
    \includegraphics[width=\columnwidth,keepaspectratio]{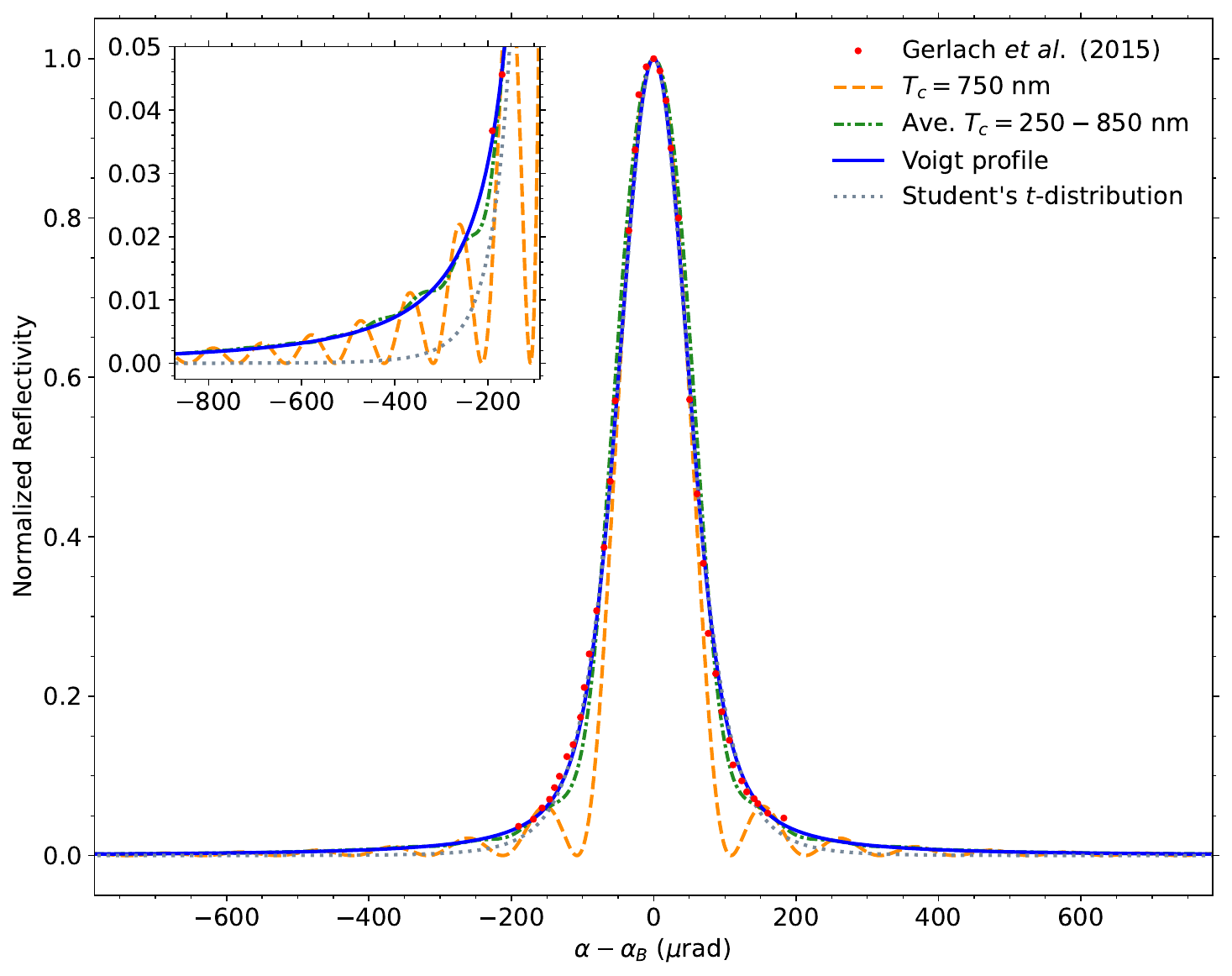}
    \caption{A comparison of the HAPG rocking curve using a single crystal thickness (orange dashed), an uniformly-averaged curve for different thickness (green dashed-dot), and experimental measurements of the rocking curve from Ref.~\cite{Gerlach_JAC_2015} (red points). The experimental curve has a FWHM of 116~$\mu$rad. The theoretical and experimental curves are for a 8.048~keV photon. Note that the theoretical curves have been shifted down in angle by 32.3~$\mu$rad to center on zero arising from dynamical diffraction theory~\cite{zachariasen1994theory}. Also shown are fits to the experimental data using a Voigt profile (blue solid) and Student's $t$-distribution (grey dotted). Inset: Focused plot on the wings of the curves. The scale of both axes is shared with the main plot.}
    \label{fig:ExpRC}
\end{figure}

\begin{figure*}
    \centering
    \includegraphics[width=\textwidth,keepaspectratio]{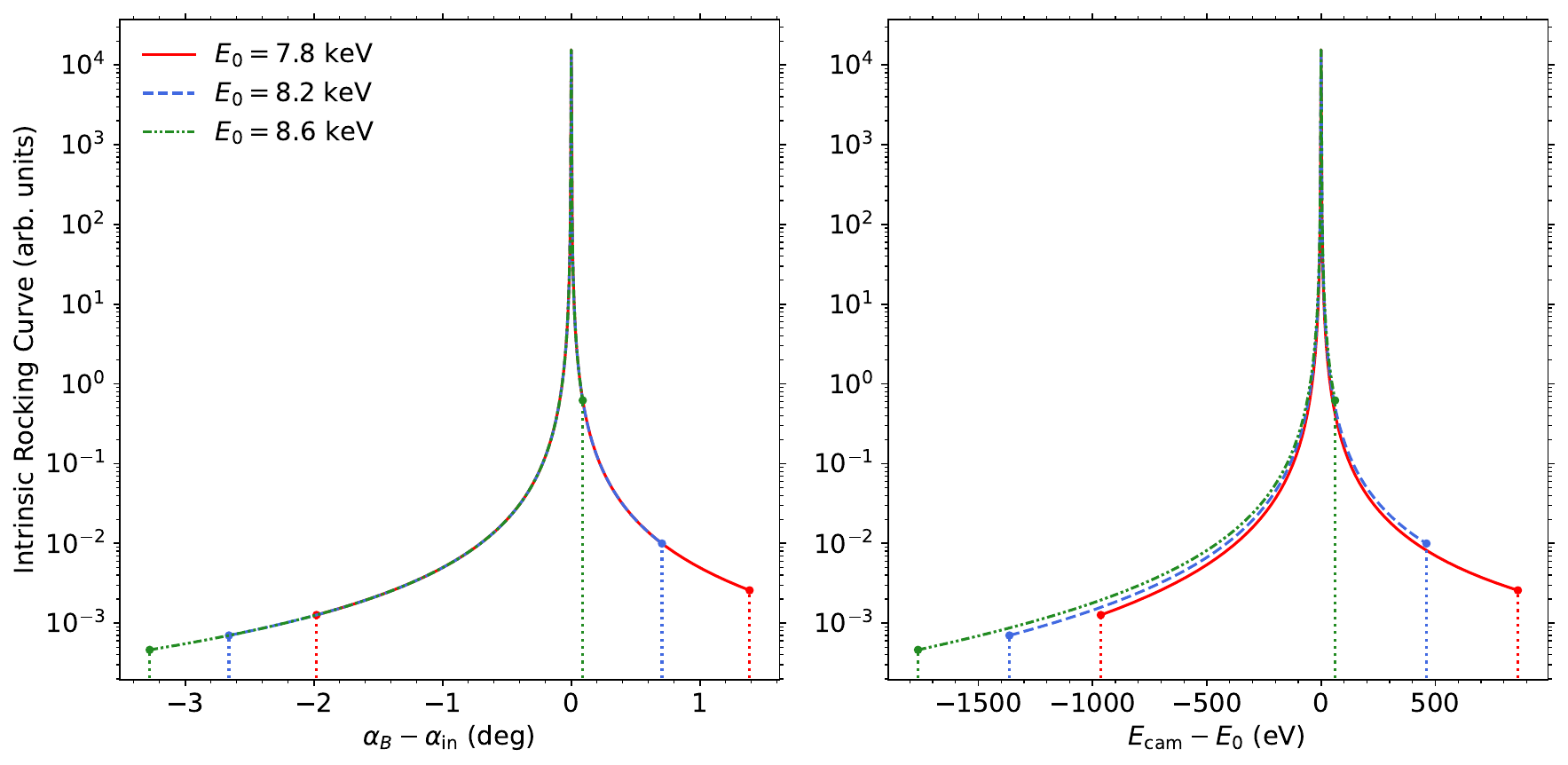}
    \caption{A Voigt profile intrinsic rocking curve in terms of the angle of incidence on the crystal (left) and the energy on the detector (right). The vertical dotted lines indicate cut-offs due to the edges of the crystal. The FWHM of the IRC is 55.7~$\mu$rad, with a Gaussian FWHM of 50.4~$\mu$rad and a Lorentzian FWHM of 9.48~$\mu$rad.
    } 
    \label{fig:IRConCam}
\end{figure*}

So far, the broadening effects considered have only shown broadening to higher photon energies, yet there is also evidently broadening towards lower photon energies~\cite{Preston_JoI_2020}. This broadening comes from the intrinsic rocking curve (IRC) of the crystallites~\cite{zachariasen1994theory}. The IRC, $W_{\rm RC}(\alpha_B - \alpha_{\rm in}, \Gamma)$, allows for diffraction to occur even when the Bragg condition is not satisfied, and it is a property of the crystallites themselves.
In general, it is a very narrow function. Indeed, for HAPG it is approximately an order of magnitude narrower than the mosaic distribution function~\cite{Gerlach_JAC_2015}. Nevertheless, it provides a very important contribution to the crystal instrument function in that it is the only source of broadening towards lower energies.
Mechanically, this is because the mosaic broadening in isolation still requires the Bragg condition to be satisfied for a given crystallite, while the IRC allows this condition to be broken. For the IRC in isolation, the reflection rule is still followed, i.e. $\alpha_{\rm out} = \alpha_{\rm in}$, hence there is broadening in both directions (cf. Fig.~\ref{fig:IRConCam}).

In the context of temperature extraction via detailed balance, the IRC is a very important contribution as the broadening towards the downshifted part counteracts the increase in intensity in the upshifted part from mosaic and depth broadening in the deconvolution.
Furthermore, even if the full SIF is symmetric in energy, the deconvolution in the Laplace domain can still lead to a reduction in the inferred temperature. For example, if the full SIF is a single Gaussian with variance $s^2$, then its two-sided Laplace transform is $\propto \exp(s^2 \tau^2)$; hence, deconvolution with a Gaussian SIF still leads to a reduction in the inferred temperature as the minimum of the ITCF is shifted towards higher $\tau$.

The rocking curve for a crystal with a certain thickness can be calculated using dynamic diffraction theory~\cite{zachariasen1994theory}, and the IRC of the crystallites in HAPG has been measured previously~\cite{legall2006new,Gerlach_JAC_2015}. 
An example of the theoretical diffraction pattern for a graphite crystallite of thickness $T_c = 750$~nm is shown in Fig.~\ref{fig:ExpRC} alongside the measurement of the IRC extracted from Ref.~\cite{Gerlach_JAC_2015}.
Evidently, the oscillations in the theoretical curve for a single thickness are absent in the wings of the experimental measurement (including Ref.~\cite{legall2006new}), and indeed no single thickness reproduces the experimental data. This may be due to the fact that the HAPG crystallites are not a single thickness, but have a distribution of sizes~\cite{2009_Legall_Efficient,2012_Zastrau_Focal}. One would therefore expect the rocking curve measurement to measure the average curve over a range of thicknesses. The rocking curve when averaged over uniformly-distributed different crystal thicknesses shows better agreement with the experimental data, but it is still not perfect. As the distribution of the crystallite thicknesses is likely to depend on the conditions in which the crystal is grown, the specific intrinsic rocking curve is likely to vary for each crystal. Indeed, it was found here that when trying to compare the fitted average function in Fig.~\ref{fig:ExpRC} to the experimental data presented later, the wings of the IRC were greatly overestimated.

The experimental measurement of the IRC, as well as the averaged theoretical curve, imply that the IRC has a functional form with tails between that of a Lorentzian and a Gaussian -- the former being unable to capture the shape near the peak of the IRC, while the latter decays in the tails far too quickly.
We note that a recent ray tracing treatment of mosaic crystals considered a rectangular function for the IRC~\cite{Smid_CPC_2021}; however, we found this function could not reproduce extended tails seen in the scattering data as, like the Gaussian, it decays far too quickly in the tails.
Natural choices then for fitting the entire distribution are those which lie between a Gaussian and Lorentzian, such as
a Voigt profile (the convolution of a Gaussian and Lorentzian) and the Student's $t$-distribution.
The $t$-distribution is essentially a generalization of the Lorentzian profile to higher powers and connects between the Gaussian and Lorentzian distributions in parameter limits. However, for $|\alpha - \alpha_B|>0.008^\circ$, the fitted $t$-distribution in Fig.~\ref{fig:ExpRC} begins to decay faster than the experimental and theoretical curves. Indeed, it was found that the $t$-distribution tended to underpredict the far wings of the broadening towards lower energies in quasi-elastic scattering data.
The Voigt profile, also shown in Fig.~\ref{fig:ExpRC}, decays very slowly in the wings. In comparison to the theoretical curves, the Voigt profile captures the slow decay of the wings better than the $t$-distribution. Indeed, the Voigt profile was found to be able to reliably fit the wings of the quasi-elastic peak, even to large energy shifts below the peak.
We therefore choose to approximate the IRC using the Voigt profile (\textsc{scipy.special.voigt\_profile} from the SciPy package for Python~\cite{2020SciPy-NMeth}), where the Lorentzian and Gaussian FWHMs are left as free parameters in the fitting. While this choice is motivated by the need for a curve that provides good fits to the quasi-elastic peak, evidently it would be preferable to have additional measurements of the HAPG IRC to higher rocking angles so that the decay of the wings could be better modelled, either with an approximate function as done here or with diffraction theory.

Like the mosaicity, the IRC also depends on the photon energy, with photons of higher energy observing an IRC with a smaller width~\cite{Gerlach_JAC_2015,zachariasen1994theory}, though like the mosaicity the IRC width depends quite weakly on photon energy.
An additional complication is that the shape of the IRC of a crystallite also depends on the photon energy~\cite{zachariasen1994theory}. But, for the photon energies of interest here (6.90-7.7~keV, 7.8-8.7~keV), both of these dependencies are assumed to be sufficiently weak that the IRC can be approximated as a Voigt profile across the entire detection range.

In Fig.~\ref{fig:IRConCam}, an example Voigt profile IRC is plotted both in terms of the angular difference and the energy on the detector, but assuming the width is fixed.
In terms of the angle of incidence, the curves have the same shape except for the position of the cut-offs due to the crystal edges.
This is complicated when the mosaicity is introduced as the IRC now allows for the Bragg condition to be broken for the crystallites. As a result, for a given angle of incidence there are now a number of crystallites which a photon can reflect off. For the present work, at each angle of incidence on the crystal, we multiply the mosaic distribution function by the IRC to account for the probability of encountering a crystallite at a given angle from the surface normal, and the reduction in reflectivity from being away from the Bragg condition.
In terms of energy observed on the detector, we observe some differences in the IRC due to the different positions of the central Bragg angle [cf. Eq.~(\ref{eq:AngleToCam})]. In other words, even keeping the angular IRC the same for a fixed photon energy, the geometry of the setup still results in variations in the broadening.

\subsection{Model Application}

\begin{figure*}
    \centering
    \includegraphics[width=\textwidth,keepaspectratio]{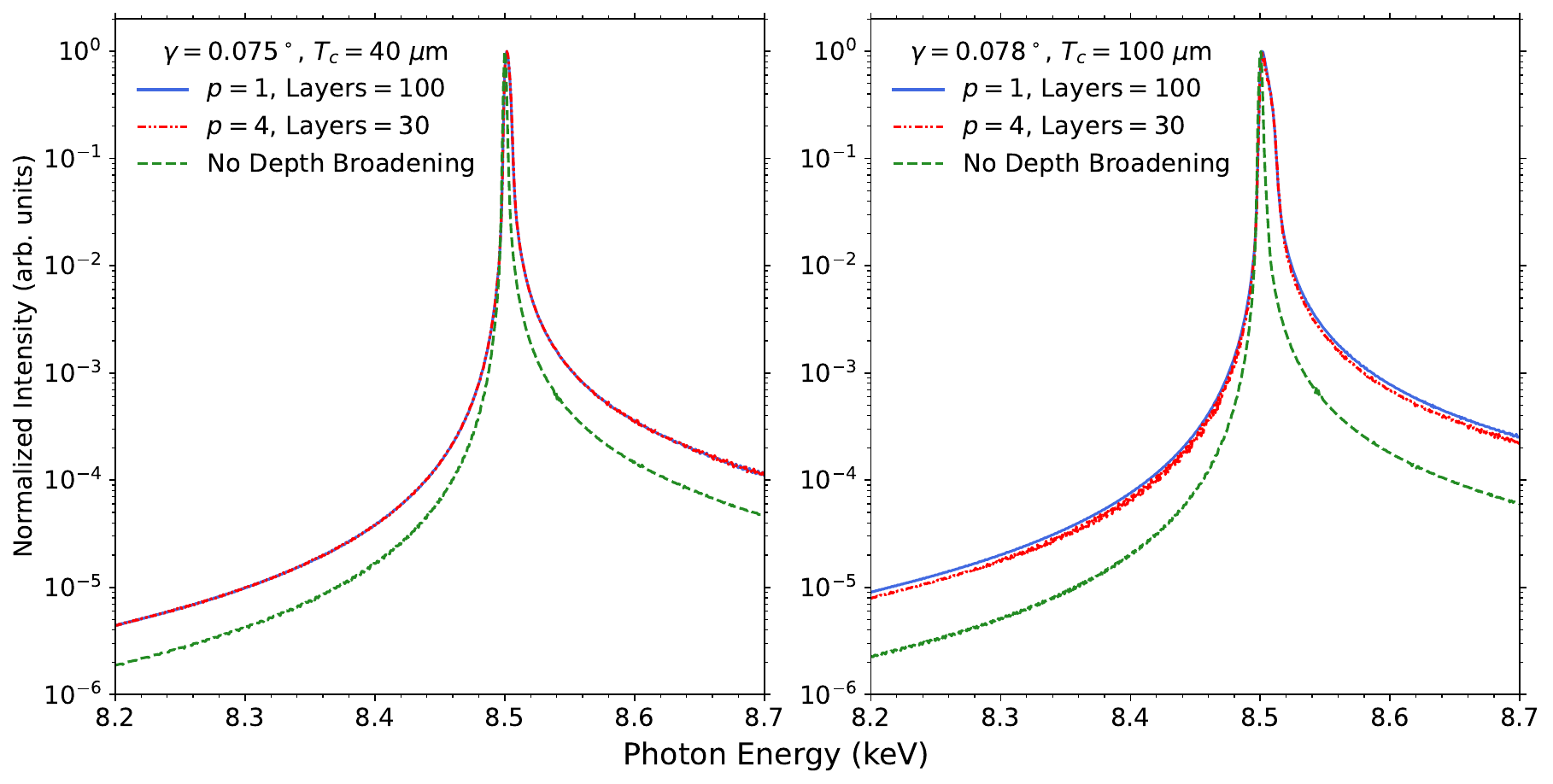}
    \caption{A comparison of the crystal instrument functions for different maximum reflection orders in 40~$\mu$m (left) and 100~$\mu$m (right) of HAPG. The green dashed-dot curve shows the IF when neglecting depth broadening.
    For the mosaicities considered here, for the 40~$\mu$m crystal shows little difference in the broadening from only $p=1$ reflections (red solid) and when including up to $p=4$ (blue dashed) reflections, while for the 100~$\mu$m crystal the multiple reflections have some noticeable contribution to the wings of the IF.}
    \label{fig:LayerEffects}
\end{figure*}

In this work, we examine the instrument broadening of mosaic HAPG crystals with parameters similar to those available at the European XFEL (EuXFEL) in Germany~\cite{Preston_JoI_2020, Zastrau_2021}. Specifically, we consider the crystals that have a radius of curvature of 80~mm, a length of 80~mm, and thicknesses of 40~$\mu$m and 100~$\mu$m.
In the present mosaic model, the source is treated as a point at the origin; the crystal is a line at a fixed distance below the source $r$, but can be varied in size and distance along the dispersion axis; and the detector is likewise a line parallel to the crystal that can be varied in height and distance from the source; see Fig.~\ref{fig:VonHamos}.
As mentioned previously, this explicitly neglects any broadening effects off the dispersion axis on the detector by taking a narrow region of interest around the focus line, and along the bent crystal because the crystal is narrow at only 30~mm.
There are also additional effects, e.g. the strain within the crystallites or from the bending process~\cite{legall2006new}, that are not explicitly accounted for in the model.
Nevertheless, this simplified treatment of the spectrometer still provides good agreement with experimental measurements.

In order to converge the shapes of the curves, 450 angles of incidence on the crystal were sampled, and the IRC was linearly sampled across 3750 angles between $\pm 1.5^\circ$.
For the $G_p$ term in Eq.~(\ref{eq:DiffRefl}), we considered up to $p=4$ and 30 reflection layers. While it is obvious that depth effects must be considered, multiple reflections within the crystal did not change the IF significantly, as can be observed in the Fig.~\ref{fig:LayerEffects}. For the 40~$\mu$m crystal, there was no observable difference between the IF whether considering a maximum order of $p=1$ or $p=4$. For the 100~$\mu$m crystal, slight differences in the IF can be observed in the far wings.

As the distribution of the crystallites in the EuXFEL's HAPG crystals has not been measured directly, we treat their distribution to be a single Lorentzian~\cite{Gerlach_JAC_2015} and the mosaicity as a free parameter to be fitted. Likewise, the Gaussian and Lorentzian components of the Voigt profile IRC are also free parameters.
These three parameters were found by performing a non-linear least squares fit of the logarithmic crystal IF to the logarithm of the quasi-elastic signal in experimental x-ray Thomson scattering data (presented in the Section~\ref{sec:results}) using the \textsc{optimize.curve\_fit} routine in the SciPy package for Python~\cite{2020SciPy-NMeth}. By using the logarithm in the fitting, we consider the relative error instead of the absolute error so that the very low intensity wings still contribute to the fitting procedure.

The procedure to calculate the crystal reflectivity is as follows: the extent and position of the crystal and detector compared to the source are defined, and the energy of the incident photon $E_0$ is also chosen. Additionally, pixels on the detector are defined for later binning the instrument function.
The crystal is then divided into a large number of sections with co-ordinates $(x_{\rm crys}, r)$ which are then associated to different angles of incidence $\alpha_{\rm in}$. 
For each $\alpha_{\rm in}$, the differential reflectivity versus $z_{\rm eff}$ is calculated using Eq.~(\ref{eq:DiffRefl}).
To determine the crystal broadening observed on the detector, the shift within the crystal is first estimated to be $\Delta x_{\rm crys} = z_{\rm eff} / \tan(\alpha_{\rm in})$, as is done in Ref.~\cite{Schlesiger_JAC_2017} -- that is, a photon is treated as travelling its full effective depth into the crystal before it encounters the crystallite it reflects off. It is also at this depth that the effect of the IRC is applied.
The final position of the photon on the detector axis is then traced and binned into the detector pixels to develop the instrument function as observed by the detector, which allows for direct comparison with experimental measurements of the instrument function.

\section{Results\label{sec:results}}

\subsection{SIF Model versus Experiment\label{sec:experiment}}

\begin{figure*}
    \centering
    \includegraphics[width=\textwidth,keepaspectratio]{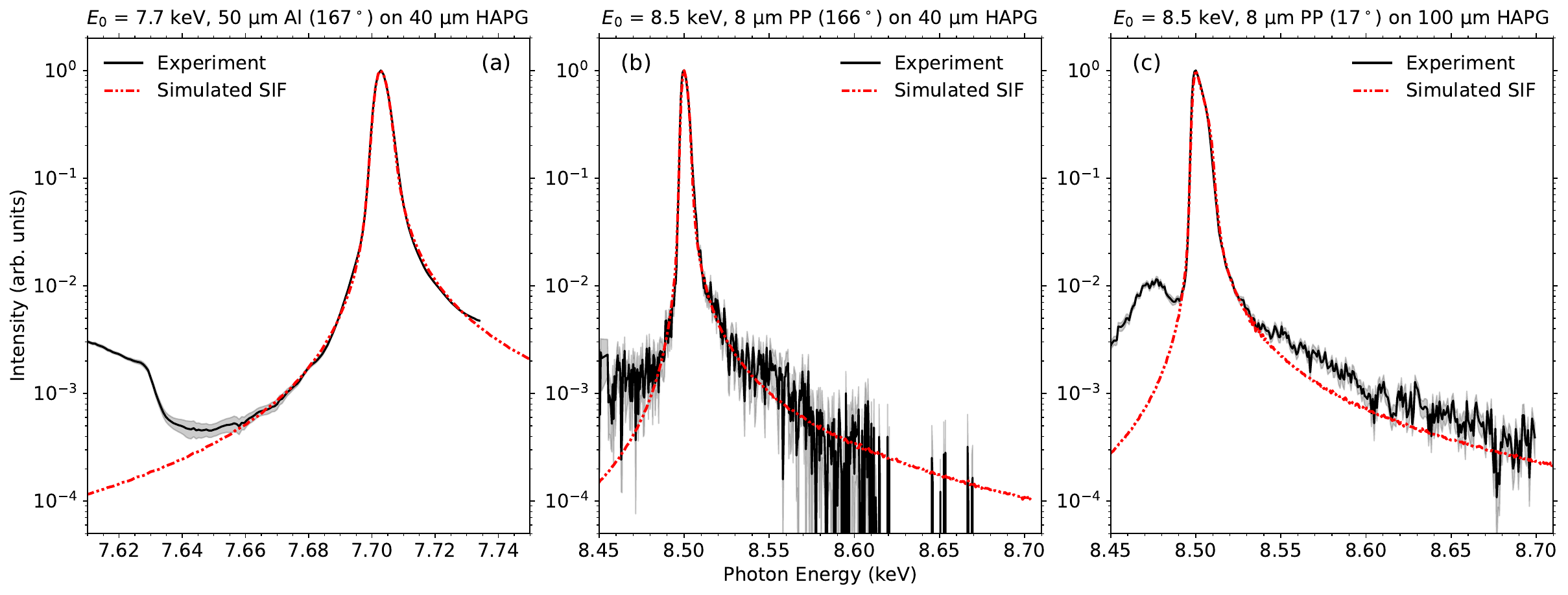}
    \caption{A comparison of the crystal IF model (red dashed) fitted to experimental measurements (black) of the quasi-elastic signal of a monochromated beam with (a) $E_0 = 7.7$~keV on 50~$\mu$m thick Al with 40~$\mu$m HAPG, (b) $E_0 = 8.5$~keV on  8~$\mu$m thick polypropylene (PP) with 40~$\mu$m HAPG, and (c) $E_0 = 8.5$~keV on 8~$\mu$m thick PP with 100~$\mu$m HAPG. The angles in the brackets are the scattering angles at which the spectra were taken. The grey shaded regions indicate the uncertainty in the experimental spectra. The theoretical SIFs are plotted on energy axes that extended beyond the experimental spectral range of the detector.
    }
    \label{fig:ModelComparison}
\end{figure*}

\begin{figure}
    \centering
    \includegraphics[width=\columnwidth,keepaspectratio]{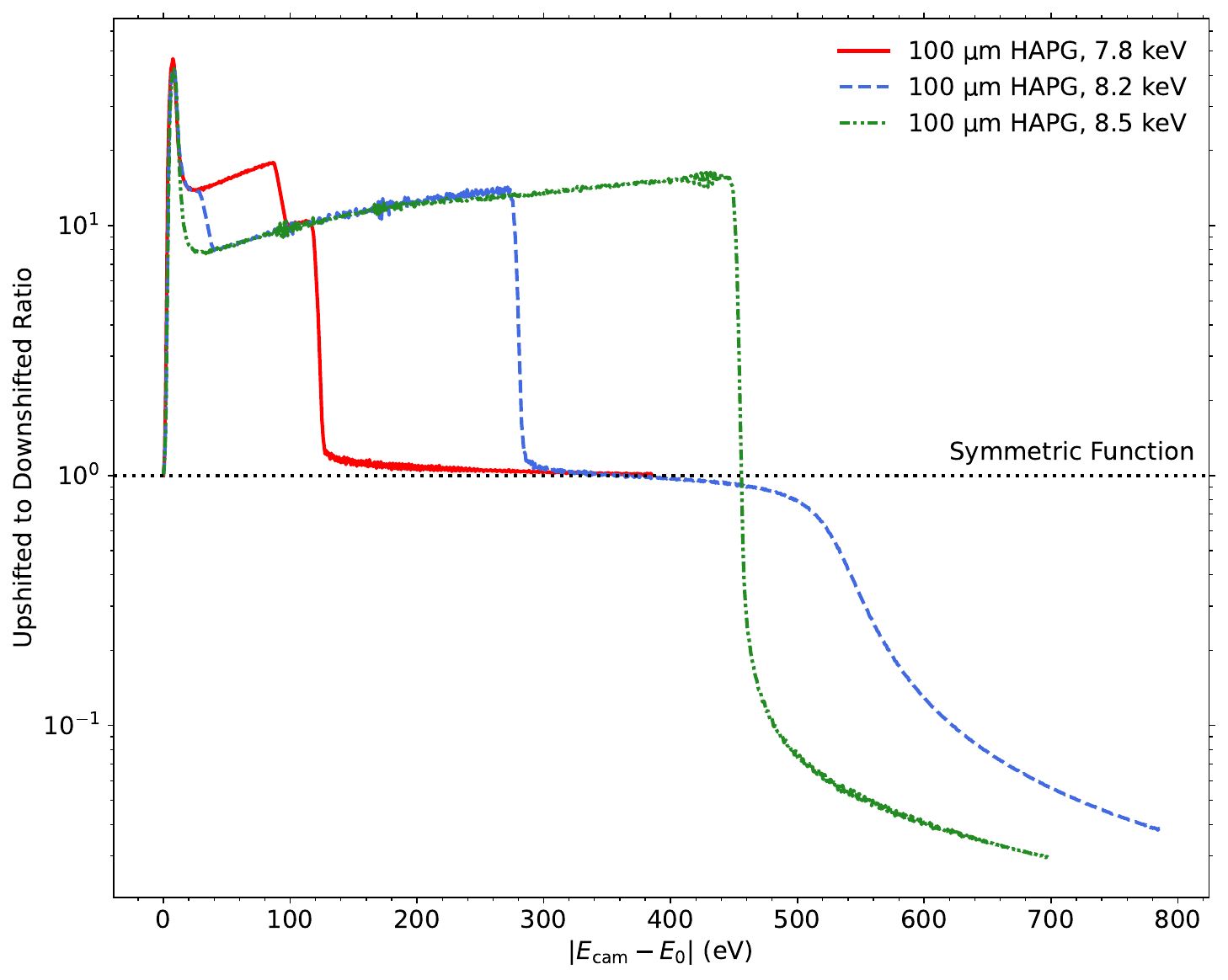}
    \caption{Ratio of the upshifted to downshifted broadening for the instrument function of the 100~$\mu$m crystal at different photon energies. The black dotted line indicates unity, i.e. the ratio for a symmetric instrument function.}
    \label{fig:SIFasymm}
\end{figure}

We start by comparing the HAPG crystal IF fit using the model on experimental measurements of the IF in order to verify the applicability of the model.
To do this, we compare the IF predicted by the model to the quasi-elastic scattering signal from XRTS measurements. Except for very high resolution setups~\cite{McBride_RSI_2018,Wollenweber_RSI_2021,Descamps_JSR_2022}, the quasi-elastic feature is essentially a delta function and so it reproduces the SIF, assuming Eq.~\ref{eq:Kernel} is a convolution.
A material that produces no inelastic scattering would be ideal as it would give direct access to the SIF. In practice, we must settle with using materials that have very weak inelastic scattering in the vicinity of the quasi-elastic feature for the particular scattering vector examined. Note that this means there is inevitably some uncertainty in the IF in the vicinity of the inelastic features. Plastics such as PMMA and polypropylene (PP) have very prominent quasi-elastic features relative to the inelastic scattering, while Al in backscattering geometry has inelastic features that are very far from the quasi-elastic signal. In combination with a monochromated beam that allows the source contribution to the SIF to be neglected, here we will use PP and Al to measure the spectrometer instrument function for experiments that were performed at the EuXFEL.

The experimental data presented in this section was collected as part of two different experiments at the EuXFEL. While the primary goals of the experiments were not to measure the instrument functions of the HAPG crystals, the experiments nevertheless provided high quality data of the quasi-elastic feature. In both cases, the FEL beam was self-seeded and then passed through a four-bounce Si (111) monochromator with an acceptance range of 0.8~eV. The monochromator has been shown to result in beams with FWHM $\sim 500$~meV~\cite{gawne2024ultrahigh}. The broadening of the signal by the beam is therefore negligible, which was verified by broadening the modelled IF by a 500~meV and not observing any difference in the spectrum.
Additionally, in both experiments the spot sizes on the targets were 10~$\mu$m. In von H\'amos geometry, there is a one-to-one imaging of the spot on the detector, so the finite spot size broadening in the dispersive direction is smaller than the pixel size of the detectors used. The broadening by these small spot sizes was therefore also found to be negligible.
We may therefore focus on the broadening by the crystal itself.
In all cases, the XRTS signal was collected in ambient conditions -- there is therefore no upshfited part to the DSF allowing for detailed examination of the broadening up to high energies above the quasi-elastic signal.
The thickness of the HAPG crystals of interest have a standard deviation of 2~$\mu$m, so the broadening due to crystal surface roughness is negligible and not included in the model.
Finally, the samples were thin, so that broadening due to scattering from different points across the sample depth were found to be negligible.

The first measurement we compare against is on 50~$\mu$m Al at a scattering angle of $167^\circ$ and a photon energy of 7.7~keV, plotted in Fig.~\ref{fig:ModelComparison}~(a). The spectrum was collected using a 40~$\mu$m HAPG crystal. The first inelastic feature is $\sim 72$~eV below the quasi-elastic feature, allowing for the broadening towards lower energies to be examined without much influence from the inelastic scattering.
We find good agreement between the model and the experimental data. Notably, the slow decay of the quasi-elastic peak to lower energies further supports the notion that the IRC decays in a manner described by a Voigt function.
As for the upshifted broadening, we also find good agreement in the spectral region that was measured, however the spectrum is only detected up to 34~eV above the elastic, and this limited range in the upshifted part of the spectrum makes it difficult to assess whether this tail has been accurately modelled.
Still, both the experimental data and the model suggest that the quasi-elastic signal is substantial and comparable to the detectable signal level -- $2.5 \times 10^{-4}$ weaker than the elastic peak in this dataset -- even $>100$~eV away from the elastic peak. Therefore, in order to meaningfully measure the full instrument function of the spectrometer, the elastic peak needs to be positioned relatively far from the detector edges.

The second measurement of the crystal IF is on 8~$\mu$m PP with a photon energy of 8.5~keV in both backwards (166$^\circ$) and forwards (17$^\circ$) scattering, plotted in Fig.~\ref{fig:ModelComparison}~(b) and (c), respectively. The backscattering data was collected on the 40~$\mu$m thick crystal while the forward scattering data was collected on the 100~$\mu$m thick crystal.
Like Al, PP has relatively weak inelastic scattering, although it is much closer to the quasi-elastic signal than in Al. As a result, the downshifted broadening is harder to distinguish from the elastic peak. However the main shape of the quasi-elastic peak and the upshifted wing is unaffected by the broadened inelastic scattering contributions as the inelastic scattering is orders of magnitude weaker than the elastic feature.
In this dataset the elastic scattering was positioned 150--200~eV below the high-energy edge of the detector so that the upshifted broadening can be examined. The experimental data indicates that there is substantial signal in the wings even some 200 eV away from the quasi-elastic peak.
In both spectra, we find the model is able to reproduce the main peak of the quasi-elastic feature.
For the 40~$\mu$m crystal, the model reproduces the upshifted broadening quite well. Above 8.53~keV, the spectrum becomes quite noisy, making it difficult to assess the quality of the model. Above 8.6~keV, the noise level of the data is reached, which explains the gaps in the spectrum as the data is collected in the single photon regime. More shots would be required to unambiguously assess the quality of the model in this region.

\begin{table*}
    \begin{center}
    \begin{tabular}{c || c | c || c | c | c || c}
     Fig.~\ref{fig:ModelComparison} Label & $T_c$ ($\mu$m)  & $E_0$ (keV)  & $\gamma$ (deg.) & $F_L$ ($\mu$rad) & $F_G$ ($\mu$rad) & $F_V$ ($\mu$rad) \\
     \hline
     (a) & 40  & 7.7  & 0.067 & $23.4$ & $83.8$ & $97.0$ \\ 
     (b) & 40  & 8.5  & 0.075 & $9.42$ & $50.4$ & $55.7$ \\  
     (c) & 100 & 8.5  & 0.078 & $9.77$ & $61.1$ & $66.5$ \\
    \end{tabular}
    \end{center}
    \caption{Fit parameters for the $\gamma$, and FWHM for the Lorentzian $F_L$ and Gaussian $F_G$ components of the IRC for the simulated instrument functions shown in Fig.~\ref{fig:ModelComparison}. Also given is the FWHM of the Voigt profile $F_V$ for these $F_L$ and $F_G$.}
    \label{tab:FitParms}
\end{table*}

The final fit parameters for the mosaicity $\gamma$, and FWHM for the Lorentzian $F_L$ and Gaussian $F_G$ components of the IRC are given in Table~\ref{tab:FitParms}.
Note that due to the depth broadening being akin to an exponential, it results in a small shift of the maximum away from the elastic peak by $\sim 1.6$~eV, which is corrected for before the modelled and experimental spectra are compared to evaluate the fit.
The three fitted mosaicities are all $\lesssim 0.1^\circ$, which is expected for HAPG~\cite{Gerlach_JAC_2015}.
In all three cases, the Gaussian component of the IRC is larger than the Lorentzian component, but the presence of the latter means the tails will be extended to high angles and at large energies shifts.

Also provided in Table~\ref{tab:FitParms} are the FWHM of the Voigt profiles $F_V$ using Kielkopf's approximation~\cite{1973_Kielkopf_JOSA}.
The two widths from the PP data are quite similar, but narrower than the measurements of the IRC on Al. Measurements of the IRC presented in Ref.~\cite{Gerlach_JAC_2015} suggest that, at the two photon energies examined here, the IRC of HAPG would be $\sim 104\pm5$~$\mu$rad. The PP measurements are substantially narrower than this, and fixing the IRC width to 104~$\mu$rad leads to an overestimation of the main peak of the quasi-elastic feature.

In all three cases, we find that the the instrument functions are highly asymmetric, as is shown in Fig.~\ref{fig:SIFasymm}. Near to the central photon energy, the broadening is strongly asymmetric towards the upshifted side of the spectrum. Once the mosaicity drops off as the edges of the crystal are reached at very high energy shifts, the broadening is then strongly asymmetric towards the downshifted side of the spectrum. However, the intensity is so low at such large energy shifts it may not be detectable. Additionally, such large energy shifts may also fall outside the spectral range of the detector.
In general then, the bulk of the broadening will be towards higher energies.
In the context of XRTS, where the ratio of the upshifted to downshifted intensity is related via the detailed balance relation, the asymmetry of the SIF will lead to an enhancement of the upshifted side of the spectrum. If a symmetric SIF is then used to infer temperature in forward modelling, or in the deconvolution step of in the imaginary-time domain, then detailed balance a higher temperature may be inferred in order to compensate for the asymmetry of the actual SIF.

\subsection{Convergence of the Model-free Analysis}

\begin{figure*}
    \centering
    \includegraphics[width=\textwidth,keepaspectratio]{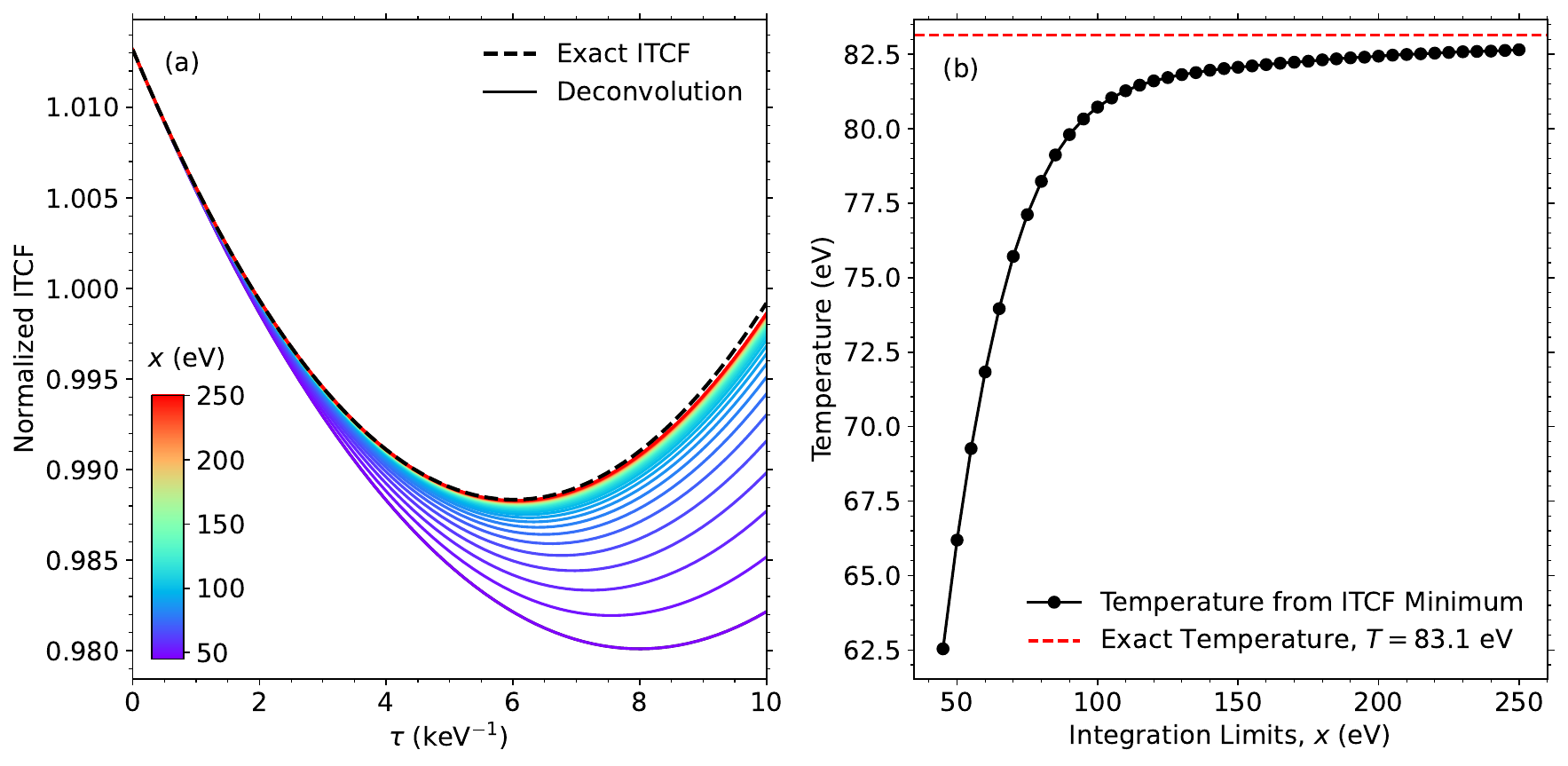}
    \caption{Convergence of the (a) the ITCF and (b) the inferred temperature with respect to the integration limits $x$ in Eq.~(\ref{eq:2sided}) in forward scattering (see red dashed curve in Fig.~\ref{fig:Convolution}~(a)). The dashed curves in each plot represents the exact result from the Laplace transform of the calculated DSF. The solid curves in the plot~(a) are are the ITCF for the different $x$, indicated by the colour of the curve. The black points in plot~(b) are the temperatures inferred from the minima of these ITCF curves.
    }
    \label{fig:ITCFconv}
\end{figure*}

In the ITCF method, the deconvolution of the SIF from the XRTS spectrum is achieved using the convolution theorem for the two-sided Laplace transform; cf. Eq.~(\ref{eq:ITCF}). More explicitly, the ITCF is calculated by the following equation~\cite{Dornheim_T2_2022}:
\begin{equation}
    \label{eq:2sided}
    \mathcal{F}(\bm{q}, \tau) = \frac{\mathcal{L}[I]}{\mathcal{L}[\xi]} = \frac{\int_{-x}^{x} I(\omega) e^{-\hbar \omega \tau} \,d\omega }{\int_{-x}^{x} \xi(\omega) e^{-\hbar \omega \tau} \, d\omega}    \, ,
\end{equation}
where the two-sided Laplace transforms in the numerator and denominator are defined when the integration limits $x \rightarrow \infty$, and we assume the effect of the SIF maybe accurately approximated as a convolution.
In practice, as the spectral window is finite, the ITCF and derived quantities such as temperature are determined by their convergence with respect to $x$~\cite{Dornheim2022-fh,Dornheim_T2_2022}. For convergence to be achieved, the spectral range examined must contain a sufficient portion of the spectrum and the SIF.

The mosaic crystal IF has been observed here to be extremely extended, predominantly because of the Lorentzian components of the mosaicity and the IRC. This raises two potential issues for the model-free analysis.
First, the two-sided Laplace transform of a Lorentzian is not finite, and so the integrals do not converge with increasing $x$. It is not immediately clear then that the ITCF should converge when performing a deconvolution in the Laplace domain to remove a SIF with Lorentzian contributions.
Second, if the SIF is very extended then the spectral range $[-x,x]$ that needs to be integrated over will also be large in order to capture a sufficient portion of the IF. If $x$ needs to be very large to observe convergence, then it may fall outside the spectral range of the spectrometer before convergence is observed. Even if it does not, the exponential enhancement of noise on the upshifted side of the spectrum eventually leads to instability in the calculation of the ITCF.
It is therefore pertinent to check that the ITCF can be converged within a reasonable spectral window.

In Fig.~\ref{fig:ITCFconv}, the convergence of the ITCF and the inferred temperature are plotted with respect to increasing integration limits $x$.
First, it is clear that the ITCF does show convergence behaviour despite the Lorentzian contributions, which is promising in the context of removing a generic SIF.
Second, within a very reasonable spectral range of $x < 200$~eV, the temperature shows convergence towards the actual temperature of the system.
Third, we note that the ITCF converges rapidly for lower values of $\tau$ with respect to increasing $x$. This is promising for the determination of the gradients of the ITCF at $\tau=0$, which physically correspond to the frequency moments of the DSF~\cite{Dornheim_PRB_2023,dornheim2023xray,Dornheim_review}.

We therefore find that, in the case that the SIF can be treated as a convolution, the slow-decaying IF of a mosaic crystal can be reliably removed from the XRTS spectrum within a reasonable spectral window. Furthermore, convergences can be observed with respect to the integration range in this region, allowing for the determination of physical properties of the system from the ITCF.

\subsection{Source-and-Instrument Function as a Convolution\label{sec:Convolution}}

\begin{figure*}
    \centering
    \includegraphics[width=\textwidth,keepaspectratio]{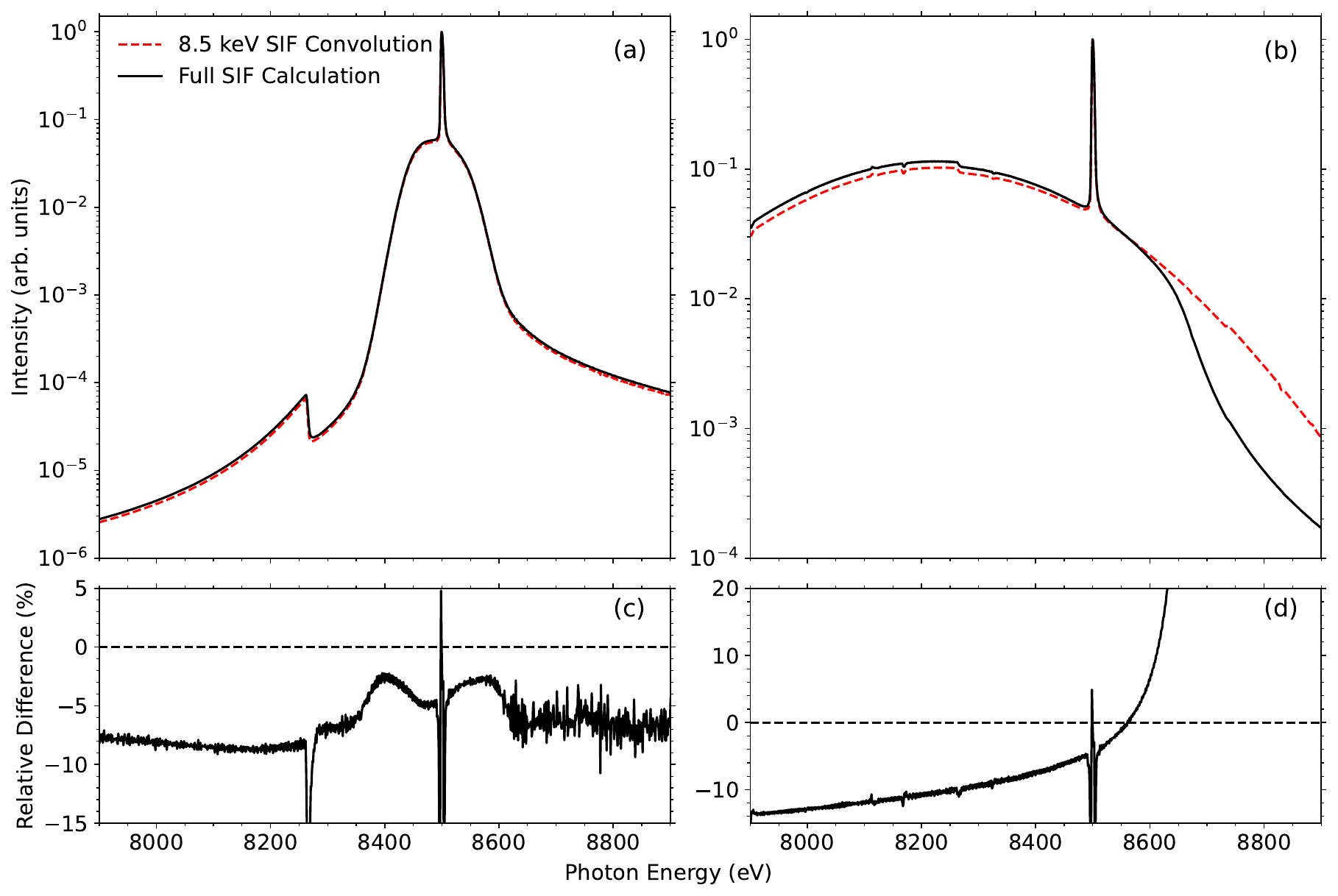}
    \caption{Top: A comparison of simulated XRTS spectra at two scattering angles -- (a) 17$^\circ$ and (b) 166$^\circ$ -- for the full SIF calculation (black solid) versus a convolution using the SIF for an 8.5~keV photon energy incident on the 40~$\mu$m crystal (red dashed). The DSF was calculated for a CH plasma at a temperature $T = 83.1$~eV, mean ionization state $Z=3.51$, and mass density $\rho=1.2$~g/cm$^3$ using MCSS. Bottom: Relative difference between the convolution and full SIF calculations.}
    \label{fig:Convolution}
\end{figure*}

\begin{figure}
    \centering
    \includegraphics[width=\columnwidth,keepaspectratio]{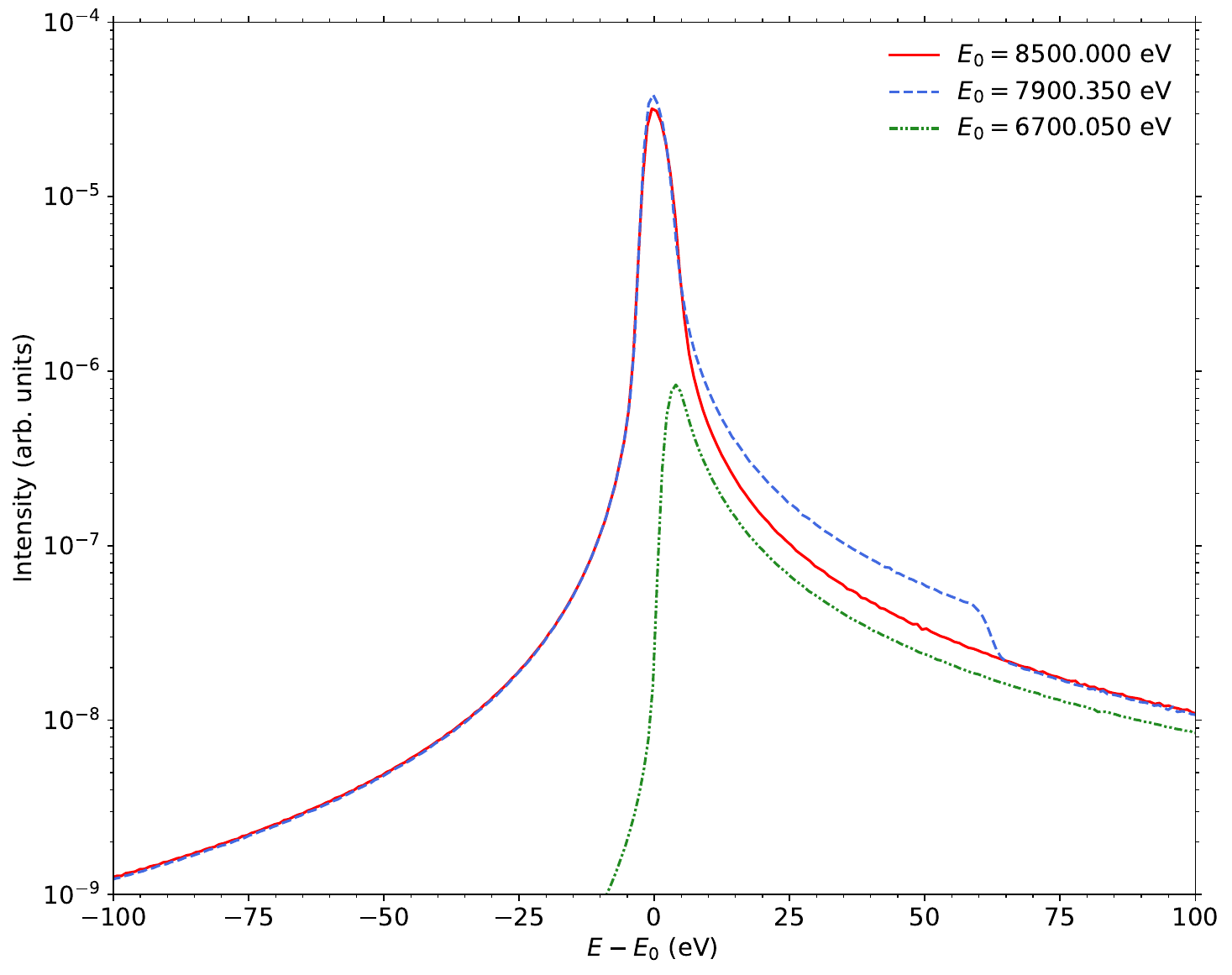}
    \caption{Comparison of the crystal IF for the 40~$\mu$m crystal (cf. Fig.~\ref{fig:ModelComparison}~(b)) at different photon energies.}
    \label{fig:SIFrange}
\end{figure}

As previously discussed, the crystal IF does not apply as a convolution to the spectrum, but it is instead a kernel with the functional form of the IF depending on the incident photon energy; cf. Eq.~(\ref{eq:Kernel}). However, a measurement of the SIF at multiple energies is infeasible for most experiments, and a convolution is more convenient for forward calculations rather than trying to model the full SIF across the entire spectral range.
As a result, in forward modelling, the SIF is generally applied as a convolution to the DSF.
Furthermore, underpinning the ITCF analysis is the idea that the effect of the SIF can be removed via the convolution theorem in the two-sided Laplace domain; cf. Eq.~(\ref{eq:ITCF}).
The alternative of solving the kernel equation to remove the effect of the SIF is a very challenging problem.
However, in both cases, so long as the SIF does not vary in energy substantially across the spectral range of the spectrometer, then its effect is essentially a convolution.

In Fig.~\ref{fig:Convolution}, we compare calculations of simulated XRTS spectra using the convolution and for a full kernel calculation.
The simulated spectra are for a CH plasma at a temperature $T = 83.1$~eV, mean ionization state $Z=3.51$, and mass density $\rho=1.2$~g/cm$^3$ calculated using the multi-component scattering simulation (MCSS) code~\cite{chapman2015probing}, with the DSF probed with a monochromated 8.5~keV beam and measured at scattering angles of 17$^\circ$ and 166$^\circ$. In the case of the convolution calculation, the convolution of the DSF with the SIF performed externally to produce the final XRTS spectrum. 
To estimate the energy dependence of the mosaicity and IRC, we use linear fits to the data reported Ref.~\cite{Gerlach_JAC_2015}, maintaining the gradients but shifting the values to match the mosaic spread and IRC widths we report in Table~\ref{tab:FitParms}~(b). 

For the 40~$\mu$m crystal, the forward scattering spectra are visually extremely similar to each other, with differences only noticeable far from the elastic where the intensity is already very low. Examining the relative difference between the two calculations shows that they are generally the same to within $\sim 5$~\% near the elastic, and around to 5-10~\% at larger energy losses. The shape of the K-edge shows differences of almost 20~\% as it is a very sharp feature that will strongly reflect the shape of the SIF.
Regardless, the absolute difference between the kernel calculation and the convolution are very small, so that the results of forward fitting and model-free approaches are likely to be quite similar.

In backwards scattering, the differences between the full kernel calculation and convolution become more apparent, with the relative difference increasing further from the quasi-elastic feature.
This is because, compared to forward scattering, there is substantial signal at energy losses far from the quasi-elastic feature in backwards scattering which will contribute to the differences in shape across the spectrum.
Additionally, there is a notable drop in intensity above 8.6~keV -- this is because the crystal position was set only to measure up to $\sim 100$~eV above the elastic (cf. Fig.~\ref{fig:ModelComparison}~(b)), so it is perhaps not surprising to see substantial differences between the two approaches. Nevertheless, this highlights the need to ensure spectrometer geometry still allows the full intensity of the signal to be captured within the spectral region of interest, as a sharp drop off such as this would certainly lead to different conclusions from the actual conditions.
Furthermore, the relation in position of the source, crystal, and detector are important to the specific shape of the spectrometer IF.

Of course, a number of materials have important features -- such as their ionization edges -- that require very large spectral windows to capture, and at some point within this spectral range the SIF will be substantially different, particularly due to the drop off in reflectivity if the crystal position is not optimized for a given photon energy. For example, Fig.~\ref{fig:SIFrange} shows an IF for a fixed crystal position for different photon energies. While there are notable differences for a 7.9~keV and 8.5~keV photon energies (both of which were intended to fall in the spectral range), the 6.7~keV photon has a substantially different SIF as it is not efficiently reflected by the crystal for this position. This effect will be mitigated to some extent by moving the crystal, though this would change the SIF for all photon energies in the spectrum as the SIF is geometry-dependent, and there are still differences that emerge because of the energy dependence of the mosaicity and IRC.

We conclude then that within a sufficiently limited spectral range, the crystal IF may be approximated as a convolution, though the applicable range and error will be dependent on the scattering and spectrometer geometry.

\subsection{Temperature via Detailed Balance\label{sec:DetailedBalance}}

\begin{figure}
    \centering
    \includegraphics[width=\columnwidth,keepaspectratio]{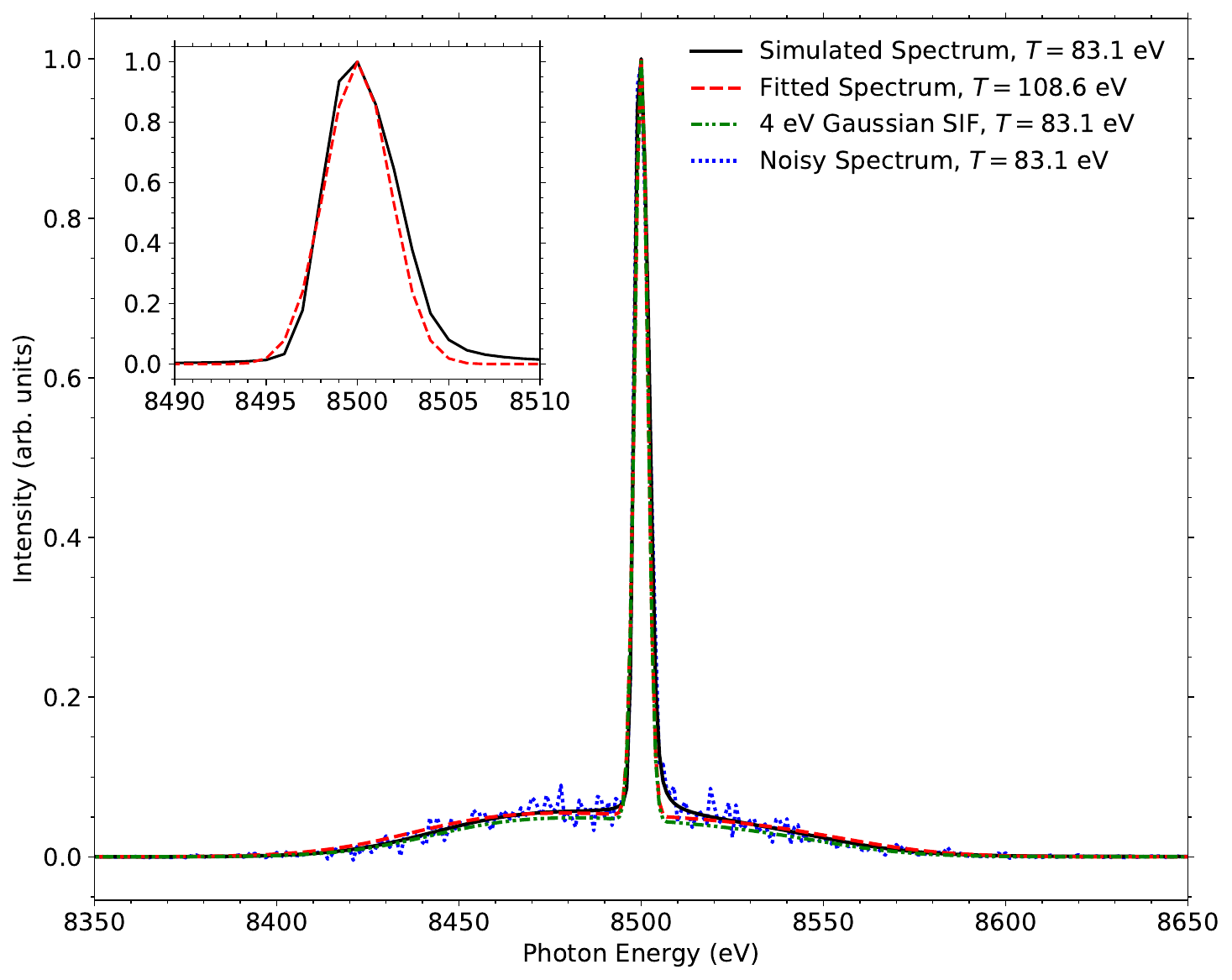}
    \caption{Comparison between a simulated experimental spectrum (red-dashed curve in Fig.~\ref{fig:Convolution}~(a)) of CH at 83.1~eV using the 40~$\mu$m crystal SIF (black solid), and a fit to this spectrum but with a 4~eV FWHM Gaussian as the estimated SIF (red dashed); a comparison of these two SIFs is shown in the inset. Also plotted are the DSF for the original black solid curve convolved with the 4~eV Gaussian (green dot-dashed), and the original solid black curve with Gaussian noise [cf. Eq.~(\ref{eq:NoisySpec})] with standard deviation $s = 0.05$ applied (blue dotted).
    }
    \label{fig:TempComp}
\end{figure}

\begin{figure*}
    \centering
    \includegraphics[width=\textwidth,keepaspectratio]{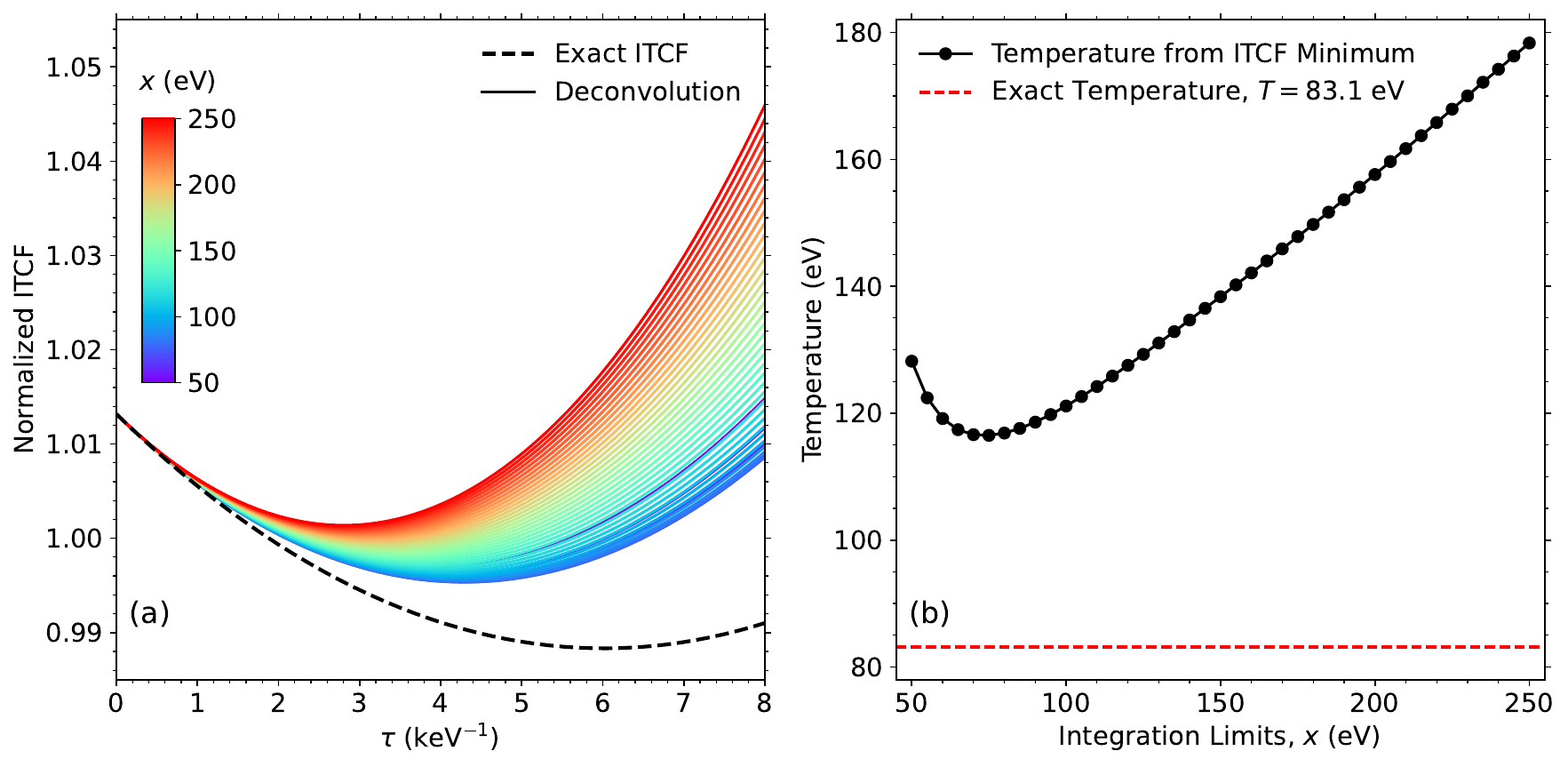}
    \caption{Convergence plots of (a) the ITCF and (b) the inferred temperature of the black curve in Fig.~\ref{fig:TempComp} when deconvolving a 4~eV FWHM Gaussian. Neither the ITCF nor the temperature show convergence to the correct temperature when using the Gaussian SIF. If the full mosaic SIF is used, both the ITCF and temperature converge to the exact the value, as shown in Fig.~\ref{fig:ITCFconv}.
    }
    \label{fig:ITCFconv2}
\end{figure*}

Finally, we consider the effect of ignoring the asymmetry in the crystal IF on plasma conditions inferred from a XRTS spectrum in experiment. In Fig.~\ref{fig:TempComp}, the simulated spectra is the same as the red-dashed curve in Fig.~\ref{fig:Convolution}~(a).
We consider here the effect of the SIF as a convolution to isolate the effect of the asymmetry in a single SIF, noting that the relative differences plotted in Fig.~\ref{fig:Convolution} indicates the full SIF can show further asymmetry between the upshifted and downshifted sides of the spectrum versus a simple convolution.

The quasi-elastic feature of this spectrum visually does not show any obvious asymmetry, and resembles a Gausian with a 4~eV FWHM. One may therefore infer that this is a reasonable estimate of the IF, as shown in the inset. 
The red dashed curve in Fig.~\ref{fig:TempComp} shows an attempt to fit to the original spectrum using the 4~eV Gaussian, with $Z$ and $\rho$ fixed to their original values.
The best fit produced requires a higher temperature of 108.6~eV, however there are very clearly differences between the two spectra that show the fitting is not perfect. Allowing the other two parameters to be free may result in some improvement.
We also plot in this figure the original spectrum with Gaussian noise applied to approximate an experimental spectrum~\cite{sheffield2010plasma,Dornheim_T2_2022} :
\begin{equation}
    \label{eq:NoisySpec}
    I_{\rm noise}(E) = I(E) + \zeta(\omega; s) \sqrt{I(E)} \, ,
\end{equation}
where $\zeta(\omega; s)$ is a Gaussian random variable centered on zero with standard deviation $s = 0.05$.
In comparison to the noisy spectrum, the higher temperature fit visually seems to be a good fit.
Lastly, we also plot the original DSF now convolved with the 4~eV Gaussian. Again, there are notable differences compared to the original smooth spectrum, but with the noise applied the Gaussian SIF seems to give reasonable fits, even with the constraints on $Z$ and $\rho$.

The fact that all those plots give reasonable looking fits to the noisy data is not entirely surprising -- the forward modelling of XRTS spectra is a notoriously unstable inverse problem when the spectrum has noise~\cite{Kasim_PoP_2019}.
However, recently laser systems at XFEL facilities have been implemented with repetition rates comparable to the XFEL ($\sim 1-10$~Hz), such as ReLaX~\cite{ReLaX} and DiPOLE-100X~\cite{DiPOLE100} at the HED instrument at EuXFEL. It is therefore now possible to rapidly collect 1,000s--10,000s of shots of warm and hot dense matter systems produced using the laser and probed with the bright FEL beam. The photon collection statistics in these setups will therefore greatly improve, and the noise diminish. As such, these subtle differences in the fitting may become more apparent, particularly as techniques to reliably remove the background to the spectra become more developed. It is anticipated the accurate determination of the SIF will become increasingly important to the accurate forward fitting of the XRTS spectrum.

We also consider the effect of using a symmetric estimate SIF on the ITCF analysis in Fig.~\ref{fig:ITCFconv2}. The figure shows the convergence in the ITCF and temperature when deconvolving the black curve in Fig.~\ref{fig:TempComp} when using the 4~eV Gaussian SIF. Evidently, neither the ITCF nor the temperature show convergence with the integration limits. The lack of convergence is because of the Lorentzian components of the mosaic crystal IF, which as described previously will not show convergent behaviour with increasing integration range $x$; cf. Eq.~(\ref{eq:2sided}). Furthermore, because the large asymmetry of the underlying SIF is not captured by the Gaussian SIF, even if the SIF is assumed to at some point decay rapidly and thereby converge, the inferred temperature will still be greatly overestimated. In comparison, it has been observed in Fig.~\ref{fig:ITCFconv} that the application of the correct SIF does result in the ITCF and temperature converging to their exact values.

It is worth noting that considering instead the noisy spectrum does not improve the ITCF or temperature determination, unlike for the forward modelling where it is superficially beneficial. The two-sided Laplace transform is very stable to noise, and so a similar lack of convergence and temperature overestimation were observed when deconvolving the noisy spectrum with the 4~eV Gaussian.

Some care is needed in the comparison of these results from the forward modelling and the model-free analysis. It may be tempting to conclude that the forward modelling is more robust than the model-free analysis on the basis that regardless of the SIF you use, you will be able to find reasonable looking fits from a forward model. 
Meanwhile the ITCF approach requires one to have much better knowledge of the SIF in order to extract properties from an XRTS spectrum.
However, this apparent robustness of the forward modelling belies the fact that an incorrect SIF still introduces an error in derived conditions. If the spectrum has noise, this error may be hidden in the fitting, but a full Markov chain Monte Carlo (MCMC) error analysis of the forward fitting~\cite{Kasim_PoP_2019} may nevertheless reveal differences in derived conditions between an estimate symmetric SIF and the true SIF.
On the other hand, the ITCF still yields meaningful and important information -- the very fact that we do not see convergence even though it is expected implies that the estimated SIF is not good enough to attempt to extract plasma conditions.

Lastly, here we have only considered the effect of the asymmetry on the inferred temperature.
However, we anticipate that it generally has an effect on all inferred properties (including, e.g., density, ionization, etc.), which should be investigated in dedicated future works.

\section{Conclusion\label{sec:summary}}

We have presented detailed calculations of the IF of mosaic crystals. Previously, the benefit of using mosaic crystals has been between their very high integrated reflectivities versus their relatively poor resolution. However, we have demonstrated the IFs of these crystals are very asymmetric to higher energies in the geometries considered here.
The asymmetry arises due to the depth broadening and the mosaic distribution of the crystallites, both of which, when isolated, only result in broadening towards higher energies. Broadening towards lower energies is facilitated by the intrinsic rocking curve of the crystallites.
If this asymmetry is not accounted for, it has a strong impact on the inferred plasma conditions even when the IF may appear to look reasonably symmetric in a spectrum. Notably, we find the temperature inferred via the detailed balance relation for an XRTS spectrum can be substantially higher when using a symmetric function, as the DSF needs to compensate for the true IF boosting the upshifted side of the spectrum. This applies in both the forward modelling and ITCF analysis.

We also find that the instrument function is not a convolution, but is instead a kernel that depends on the photon energy in a non-trivial manner. Furthermore, the specific shape of the IF depends on the geometry of the set up, which results in even greater deviations and some notable features within the IF that vary across the detector. Nevertheless, for a sufficiently narrow spectral range the kernel integration can be approximated as a convolution, which greatly simplifies forward model fitting to XRTS spectra, and allows the model-free ITCF method to be employed with high accuracy. However, the applicable energy range is still relatively narrow and will depend on the spectrum being measured.
Furthermore, the geometry dependence of the spectrometer IF suggests that while detailed studies of the spectrometer components are very useful for understanding the behaviour of the spectrometer, a measurement of the specific IF for a given experiment is still important.

Here, we have only considered the effect of the asymmetry of the SIF on the inferred temperature. Further work is required to investigate the influence of the full SIF on the inference of other plasma properties inferred via XRTS, such as the frequency moments of the DSF~\cite{Dornheim_PRB_2023}, the plasma density, and ionization. Nevertheless, we expect the choice of SIF model will have an effect on properties inferred.

Lastly, while we have focused on the effects of the crystal IF on the XRTS spectra, it will also be important to other forms of analysis. For example, where lineshapes are used  to determine plasma conditions the instrument function will evidently have an impact on the shape of the measured spectrum. Additionally, recently reported resonant inelastic x-ray scattering (RIXS) measurements on Fe and Fe$_2$O$_3$~\cite{Forte_RIXS_2024}  show clear differences in the experimentally measured vacant density of states of the materials, but the resolution and shape of these features is constrained by the IF of the mosaic crystal used in the spectrometer.

Therefore, as an overarching conclusion, we find that it is important to take accurate measurements of the spectrometer instrument function in experiments where its shape ultimately determines the analysis.

\section*{Acknowledgments}
This work was partially supported by the Center for Advanced Systems Understanding (CASUS), financed by Germany’s Federal Ministry of Education and Research (BMBF) and the Saxon state government out of the State budget approved by the Saxon State Parliament.
This work has received funding from the European Union's Just Transition Fund (JTF) within the project \emph{R\"ontgenlaser-Optimierung der Laserfusion} (ROLF), contract number 5086999001, co-financed by the Saxon state government out of the State budget approved by the Saxon State Parliament.
This work has received funding from the European Research Council (ERC) under the European Union’s Horizon 2022 research and innovation programme
(Grant agreement No. 101076233, ``PREXTREME''). 
Views and opinions expressed are however those of the authors only and do not necessarily reflect those of the European Union or the European Research Council Executive Agency. Neither the European Union nor the granting authority can be held responsible for them.

We acknowledge the European XFEL in Schenefeld, Germany, for provision of X-ray free-electron laser beamtime at the Scientific Instrument HED (High Energy Density Science) under proposal numbers 3777 and 5690 and would like to thank the staff for their assistance. The authors are grateful to the HIBEF user consortium for the provision of instrumentation and staff that enabled these experiments. The original datasets can be found here and are available upon reasonable request: doi:10.22003/XFEL.EU-DATA-003777-00, doi:10.22003/XFEL.EU-DATA-005690-00.

\section*{Author Declarations}
The authors have no conflicts of interest to disclose.
T.G., T.D., H.B. performed the analysis, with input from Z.M., J.V.. T.G. developed the code for the model, with contributions form L.B.F. and M.J.M..
T.G., H.B., K.A., C.B., V.B., E.B., D.B., A.C., A.D., S.G., N.J.H., M.-L.H, P.H., H.H., O.S.H., Z.K., A.L., B.L., J.L., M.M., W.M., M.M. Z.A.M., M.N., J.-P.N, P.N., A.P., C.Q., L.R., J.R., T.T., L.W., U.Z., D.K., T.R.P., T.D. were involved in the collection of the experimental data.
T.R.P. was the main proposer and principal investigator of experiment proposal 3777. D.K. was the main proposer and principal investigator of experiment proposal 5690.
T.G. wrote substantial parts of the manuscript, with contribution from T.D. and T.R.P. All authors were provided the opportunity to review the manuscript.

\section*{Data Availability}
The data that support the findings of this study are available from the corresponding author upon reasonable request.


\bibliography{bibliography}
\end{document}